\documentclass[
 reprint,
 amsmath,amssymb,
 aps,
]{revtex4-2}

\usepackage{graphicx}
\usepackage{adjustbox}
\usepackage{multirow}
\usepackage{array} 
\usepackage{dcolumn}
\usepackage{bm}
\usepackage{xcolor}

\begin{document}

\preprint{APS/123-QED}

\title{Resource and population dynamics in an agent-environment interaction model}

\author{Gaston Briozzo$^{1, 2, 3}$}
\author{Gustavo J. Sibona$^{1, 2}$}
\author{Fernando Peruani$^{3}$}
\affiliation{$[1]$ Facultad de Matemática, Astronomía, Física y Computación (FaMAFyC), Universidad Nacional de Córdoba (UNC), Ciudad Universitaria (5000), Córdoba, Argentina,}
\affiliation{$[2]$ Instituto de Física Enrique Gaviola (IFEG), Consejo Nacional de Investigaciones Científicas y Técnicas (CONICET), Ciudad Universitaria (5000), Córdoba, Argentina,}
\affiliation{$[3]$ Laboratoire de Physique Théorique et Modélisation (LPTM) (UMR 8089), Cergy Paris Université (CYU), 2 avenue A. Chauvin, Cergy-Pontoise cedex (95302), France\\}

\date{\today}

\begin{abstract}

In any ecosystem, the conditions of the environment and the characteristics of the species that inhabit it are entangled, co-evolving in space and time.
We introduce a model that couples active agents with a dynamic environment,  interpreted as a nutrient source. Agents are persistent random walkers that gather food from the environment and store it in an inner energy depot. This energy is used for self-propulsion, metabolic expenses, and reproduction.
The environment is a two-dimensional surface divided into patches, each of them producing food. Thus, population size and resource distribution become emergent properties of the system. Combining simulations and analytical framework to analyze limiting cases, we show that the system exhibits distinct  phases separating quasi-static and highly motile regimes. We observe  that, in general, population sizes are inversely proportional to the average energy per agent. Furthermore, we find that, counter-intuitively,  reduced access to resources or increased metabolic expenditure can lead to a larger population size. The proposed theoretical framework provides a link between active matter and movement ecology, allowing to investigate short vs  long-term strategies to resource exploitation and rationing, as well as sedentary vs wandering strategy. The introduced approach may serve as a tool to describe real-world ecological systems and to test environmental strategies to prevent species extinction.

\end{abstract}


\maketitle

\section{Introduction}

    In recent years, the field of movement ecology has attracted significant attention, driven by advancements in experimental tracking devices, the development of a unifying research framework, and new computational models, among other factors
    \cite{nathan2008movement, hoover2020digital, wisnoski2023scaling, hoffman2024benchmark}. Its relevance extends from understanding biological ecosystems to practical applications in conservation and harvest management.

    In nature, a remarkable variety of movement strategies has evolved across species as mechanisms for survival and development, {from run and tumble in bacteria to migration in large mammals}. 
    Most of these organisms invest a considerable amount of energy in exploring their environments in search of resources 
    \cite{charnov1976optimal, ginelli2015intermittent, lukas2023multispecies, pacher2024evidence, gomez2022intermittent}.
    At the microscopic scale, a substantial amount of energy is required to overcome stochastic fluctuations, which hinder directional movement. In this context, bacteria can be broadly categorized into those that actively navigate their environment in search of resources and those whose exploration relies on diffusive processes governed by environmental noise \cite{ liu2002biomimicry, condat2005randomly, perez2019bacteria, weady2024mechanics}. At larger scales, some animal species roam actively through the environment in search of prey, exhibiting more directed movement, while others remain in localized areas, foraging on nearby vegetation in a more {homing or random} movement \cite{darowski1988bioenergetic, bernoff2020agent, cooke2022movement, gomez2023fish}.

    Although the physical laws governing movement across scales differ fundamentally, a unifying principle persists: organisms seek to maximize resource acquisition while minimizing energy expenditure and exposure to risk \cite{mathijssen2018nutrient, li2023biological, masello2017animals, halsey2016terrestrial}.    
    The availability and spatial distribution of nutrients and resources determine which strategies—movement patterns, home range, among others—are viable for a given organism \cite{sumpter2010collective, cavagna2010scale, ryan2017competition},  
    while the organism's ability to adopt such strategies determines its survival and proliferation \cite{zhang2019insect, fryxell2008multiple,  van2013bacterial}. 
    For predators, this dynamic depends not only on prey abundance but also on capture efficiency, which in turn is influenced by both the mobility and evasiveness of the prey \cite{dickie2022resource}.
    In many cases, prey abundance is closely linked to habitat characteristics.     
    Moreover, temporal fluctuations in nutrient levels may act as a destabilizing factor in the dynamic behavior of populations interacting within an ecosystem—for example, triggering a shift (or bifurcation) from a stable state to population cycles or even extinction \cite{turchin2001availability}.
    As a result, there is substantial experimental evidence connecting movement patterns to habitat variation—ranging from bacteria in the gastrointestinal tract \cite{otte} to
    birds in forests \cite{korpach2025high}, large fish in the ocean \cite{muhling2025climate}, and mammals in the savanna \cite{keeping2014animal}. 
    The significance of these studies grows even more in light of climate change's effects on habitats, as well as the ongoing efforts in conservation and management.

    Consequently, developing a theoretical framework to study how self-propelled organisms interact with and adapt to energetically dynamic environments is essential for understanding species evolution driven by the optimization of (reproductive) fitness \cite{hallatschek2023proliferating}. To the best of our knowledge, there is no active particle model that incorporates the dynamics of an internal energy reservoir regulating not only agent motion, but also agent death and proliferation, while simultaneously coupling energy intake to an environment that is itself dynamic. Existing models have related locomotion efficiency to energy expenditure, but without accounting for population dynamics or  feedback on the environment. For example, the SET model \cite{schweitzer1998complex} considers a constant population of agents moving under the combined influence of Brownian forces and an active propulsion mechanism fueled by an internal energy reservoir. In this framework, agents harvest energy from the environment and convert it into kinetic energy, with energy supply modeled either as constant \cite{condat2002diffusion}, localized in space \cite{schweitzer1998complex}, or velocity-dependent \cite{di2014enhancement}. Yet, real ecological systems are characterized by populations that fluctuate in size, which are strongly coupled to the dynamics of environment that evolves in both space and time under the influence of the agents themselves. This coupling implies a non-trivial interplay between agent motion—which governs environmental exploration and exploitation—and population size.

    Here, we investigate this fundamental interplay in an agent system coupled to a dynamical environment that serves as a food supply for the agents. Agents consist of self-propelled persistent random walkers, that gather food from the environment, store it in an inner energy depot and transform it into kinetic energy or metabolic expense. 
    This framework enables population variability by incorporating birth and death processes:
    agents can reproduce when conditions are favorable and starve to death when resources are scarce.
    The environment is active and heterogeneous, discretized into a grid of identical patches that locally generate and store food, which can be consumed by the agents inhabiting it. 
    In Sec. \ref{AEM} we present the individual-based model (IBM), used to perform numerical simulations. We explain the active environment in Sec. \ref{Env}, the agents' dynamics in Sec. \ref{Age}, and the birth and death processes in Sec. \ref{BnD}.
    In Sec. \ref{AL}, we analyze {certain analytical limiting cases} (AL) to develop a formal description and derive mathematical expressions. In Sec. \ref{EB}, we derive expressions for agents' energy and spatial distributions, and in Sec. \ref{s:EP}, for the population size.    
    In Sec. \ref{NR}, we present the numerical results obtained from the IBM simulations and compare them with the AL results. Here, we explore the parameters and analyze the characteristics of the model. These results allow us to determine the optimal dynamics that maximize agent populations or resource utilization.
    This work provides a bridge between active matter and ecology by emphasizing environmental conditions and their relationship with populations and their behaviors.

\section{Agent-Environment Model}
\label{AEM}
    
    The model represents an ecological environment consisting of a two-dimensional space that contains both 
    a dynamical field—the nutrients (e.g. grass)—
    and a number $N(t)$ of biological motile entities (agents) that need those resources to reproduce and avoid starvation. 
    In the following, we will describe separately the components of the model.
 
\subsection{Environment {as a Nutrient Source}}
\label{Env} 
     
    The environment 
    is characterized by the dynamical \textit{food field} $f(\mathbf{x},t)$, which represents the local concentration of food resources.
    The environment is modeled as a bidimensional square plane of side $L$ with periodic boundary conditions, which is divided into $M$ identical square patches of side $L_p$.
    Thus, patch $\alpha$-th corresponds to the surface $S_\alpha$, for $\alpha\in\{1,...,M\}$.
    Food is homogeneous within each patch; 
    {therefore, any position $\mathbf{x} \in S_{\alpha(\mathbf{x})}$, has the same local density of food available for the agents, $f_{\alpha(\mathbf{x})}(t)$, where $\alpha(\mathbf{x})$ gives the patch index where $\mathbf{x}$ belongs.}
    The size and position of the patches are constant, and nutrients do not diffuse between them. Therefore, patches do not interact directly with each other.
    In the absence of agents, the inner energy of the patches evolves logistically \cite{turchin2001availability}. That is, when the food value is low, it grows exponentially with \textit{growth rate} $r$. As the energy increases, growth stops as food approaches the \textit{patch capacity} $c$. This represents the maximum amount of food that the patch can hold. 
    However, when agents are introduced, the energy of the patches decreases as it is consumed by them, being the agents \textit{intake rate} $q[f_\alpha(t)]$ {(described in the following section)} dependent on the patch's energy availability. 
    Thus, the temporal evolution of the available food is given by: 
    \begin{equation}
        \dot{f}_\alpha(t) =  r f_\alpha(t) \left[1-\frac{f_\alpha(t)}{c}\right] - n_\alpha(t) q[f_\alpha(t)],
        \label{eq:M_E}
    \end{equation}
    where $n_\alpha(t)$ is the number of agents in patch $\alpha$ at time $t$. 
    
\subsection{Agents}
\label{Age} 

    The environment is inhabited by a population of $N$ self-propelled, disk-shaped, agents of radius $R$.    
    Agent $i$ has a position $\mathbf{x}_i(t)$, corresponding to its center, and an orientation $\theta_i(t)$, corresponding to its propulsion direction. 
    Agents explore space by performing persistent random walks \cite{sibona2007evolution, granek2024colloquium} with constant \textit{active speed} $v_0$, changing direction randomly with an \textit{angular diffusion coefficient} $D_\theta$.
    Thus, the effective \textit{diffusion coefficient} {(in absence of interactions)} results in $D=v_0^2/2D_\theta$.
    Agents $i$ and $j$, at distance $r_{ij}(t)=\| \mathbf{x}_i(t) - \mathbf{x}_j(t) \|$, interact with each other through a soft-body repulsive potential $U[r_{ij}(t)]$ which penalizes overlapping {(see Appendix \ref{App:Pot} for its description)}. 
    This force only acts when they are in contact, i.e., the distance between the agents' centers is less than their diameters, $r_{ij} \leq 2R$.
    Agent $i$ gathers food from the environment at \textit{intake rate} $q[f_{\alpha(\mathbf{x}_i)}(t)]$ and stores it in its \textit{inner energy depot} $e_i(t)$ \cite{schweitzer1998complex, fieguth2022hamiltonian}.
    This energy is dissipated at a \textit{metabolic rate} $m$ due to the agent's own metabolic expenditures, and a constant power $\kappa v_0^2$ is used to self-propel, being $\kappa$ the \textit{kineticis used to self-propel. Thus, the state of theis used to self-propel. Thus, the state of the rate}.    
    Thus, the state of the $i$-th agent is given by its position $\mathbf{x}_i$, its orientation $\theta_i$ and its energy $e_i$, whose dynamics satisfy
    \begin{equation}
        \begin{split}
        \dot{\mathbf{x}}_i(t) &= v_0\hat{\theta}_i(t) - \nabla \sum_{j\neq i}U[r_{ij}(t)], \\
        \dot{\theta}_i(t)  &= \sqrt{2D_\theta} \xi_i(t), \\
        \dot{e}_i(t) &= q[f_{\alpha(\mathbf{x}_i)}(t)] - me_i(t) - \kappa v_0^2,
        \label{eq:M_A}            
        \end{split}
    \end{equation}
    where $\hat{\theta}_i=(\cos{\theta_i},\sin{\theta_i})$, and $\xi$ is a Gaussian noise such that $\langle \xi_i(t) \rangle =0$ and $\langle \xi_i(t_1)\xi_j(t_2) \rangle = \delta_i^j \delta(t_1-t_2)$.
    
    In particular, for this work we consider an agent intake rate defined by
    \begin{equation}
        q[f_{\alpha(\cdot)}] = I_c \tanh{\left[f_{\alpha(\cdot)} \frac{I_s}{I_c}  \right]}, 
        \label{eq:M_q}
    \end{equation}
    where the \textit{intake capacity} $I_c$ and the \textit{intake slope} $I_s$ are constants. 
    Note that agents located in the same patch have the same intake.
    For $f I_s/I_c<1$, $q(f)$ is an almost linear function with slope $I_s$.
    Thus, under conditions of scarcity, we have $q(f) \approx f I_s$. Meanwhile, for $f I_s/I_c>1$, the intake saturates at its maximum value $I_c$. 
    Thus, under conditions of abundance, $q(f) \approx I_c$.   
    
\subsection{Births and Deaths}
\label{BnD} 

    In order to obtain a dynamic population sensitive to the environmental condition, birth and death processes for the agents are introduced.
    This allows studying the impact of spatial dynamics on nutrient acquisition and of competition between agents for nutrients. 
    
    To model the death process we will consider that, if the energy of agent $i$, $e_i(t)$, decreases under the \textit{starvation energy} $e_s$, it will be removed from the system, i.e., it dies.
    On the other hand, to model the proliferation process, 
    if $e_i(t)$ increases above the \textit{reproduction energy} $e_r$, it will reproduce, i.e., it will duplicate by generating a new agent, being an identical copy of itself.
    When reproducing, the offspring is first placed in the same position as the original agent, then a random direction is taken and both are displaced a distance $R/2$ in opposite directions.
    Additionally, both inner energies (agent and offspring) are set to the \textit{birth energy} $e_b$. In this way, the energy cost of reproduction can be incorporated by assuming 
    \begin{equation}
        e_s<e_b<\frac{e_r}{2},
        \label{eq:M_BD_e}
    \end{equation}
    ensuring that reproduction has an energy expenditure, while agents are not left on the brink of death. 
    Therefore, population size $N(t)$ will be an emergent quantity,
    depending on the interplay between the environment and the agents.

\section{Analytical Limiting (AL) Cases}
\label{AL}

    In this section, we explore the AL and obtain analytical expressions for some special asymptotic cases, that allow us to find optimal agent characteristics that maximize the agent population.

\subsection{Energetic Balance}
\label{EB}

    The energetic equilibrium for a single patch with $n_\alpha$ agents in it can be derived from Eqs. \eqref{eq:M_E} and \eqref{eq:M_A},
    considering the above mentioned limit cases of food scarcity and food abundance.
    
    When the patch food is abundant, such that $f_\alpha I_s/I_c \gtrsim 1$, the agent intake saturates at a constant value, $q=I_c$, for which the equilibrium values are
    \begin{equation}
        \begin{split}
        f_{c}^{*}(n_\alpha) &= \frac{c}{2} \left( 1 + \sqrt{1 - \frac{n_\alpha}{n_{max}}} \right),  \\
        e_{c}^{*}(n_\alpha) &= \frac{I_c-\kappa v_0^2}{m},
        \end{split}
        \label{eq:EB_Ic_eq}
    \end{equation}
    being
    \begin{equation}
        n_{max} = \frac{1}{4}\frac{c r}{I_c}
        \label{eq:EB_Ic_n0}
    \end{equation}
    the maximum population that the patch is able to maintain.
    Note that equilibrium agent inner energy, $e_{c}^{*}$, is constant and does not depend on $n_\alpha$ or $f_{c}^{*}$.

    When the patch food is scarce, such that  $f_\alpha I_s/I_c \ll 1$, the agent intake grows linearly with the nutrient availability, $q=f_\alpha I_s$, for which the equilibrium values are
    \begin{equation}
        \begin{split}
        f_{l}^{*}(n_\alpha) &= c \left(1-\frac{n_\alpha}{n_{max}^{f}}\right),  \\
        e_{l}^{*}(n_\alpha) &= \varepsilon \left(1-\frac{n_\alpha}{n_{max}^{e}} \right),
        \end{split}
        \label{eq:EB_Is_eq}
    \end{equation}
    where we define
    \begin{equation}
        \begin{split}
        \varepsilon  &= \frac{c I_s - \kappa v_0^2}{m}, \\
        n_{max}^{f}   &= \frac{r}{I_s},  \\
        n_{max}^{e}   &= \frac{r}{I_s} \left( 1 - \frac{\kappa v_0^2}{I_sc} \right),
        \end{split}
        \label{eq:EB_Is_n0}
    \end{equation}
    being $\varepsilon$ the maximum balance energy,
    $n_{max}^{f}$ is the maximum population that the patch is able to maintain while $f_\alpha>0$,
    and $n_{max}^{e}$ is the maximum population at which agents are able to coexist while $e_i>0$.
    Note that $n_{max}^{e} \leq n_{max}^{f}$. 
    Moreover, for a variable population $n_\alpha(t)$, agents will die when they reach $e_l^*[n_\alpha(t)] = e_s$, which defines a limiting population for the system.

    From the results obtained for a single patch, we can now analyze a system composed of $M$ (constant) patches and $N(t)$ agents distributed among them,
    such that $n_{1}(t)+...+n_{M}(t)=N(t)$. 
    For the moment, as a first approach, we will consider a constant population, ignoring births and deaths: {$N(t):=N_0$}. Thus, the population in a patch may be beyond the limit values found above, while the agent energy, $e_i$, and nutrients availability, $f_\alpha$, vanish in that patch. 
    Note that the energies should satisfy $e_i\geq0$ and $f_\alpha\geq0$.
    
    Let $\mathbf{n}=\left(n_{1}, ..., n_M\right)$ be the population vector, organized such that $n_\alpha \geq n_\beta ~ \forall ~ \alpha \leq \beta$.
    Consider $M_{o}^{f}$ overpopulated patches where $n_\alpha \geq n_{max}^{f}$ and $M_{o}^{e}$ overpopulated patches where $n_\alpha \geq n_{max}^{e}$.  
    Let $\mathbf{n}_{o}^{f}=\left(n_{1}, ..., n_{M_{o}^{f}}\right)$ and $\mathbf{n}_{o}^{e}=\left(n_{1}, ..., n_{M_{o}^{e}}\right)$ be the 
    overpopulation vectors for patches ($f_\alpha=0$) and agents ($e_i=0$), respectively, such that $n_{1}+...+n_{M_{o}^{f}}=N_{o}^{f}$ and $n_{1}+...+n_{M_{o}^{e}}=N_{o}^{e}$
    (note that $N_{o}^{f} \leq N_{o}^{e}$ from Eq. \eqref{eq:EB_Is_n0}).
    Therefore, the expected mean energies values are
    \begin{equation}
        \begin{split}
        \left< f^{*} \right> &= \frac{1}{M} \sum_{\alpha=M_{o}^{f}+1}^M f^{*}(n_\alpha), \\ 
        \left< e^{*} \right> &= \frac{1}{N} \sum_{\alpha=M_{o}^{e}+1}^M e^{*}(n_\alpha) n_\alpha.
        \end{split}
        \label{eq:EB_fe_eq}
    \end{equation}
    In the limiting case of abundance, $q(f_\alpha)=I_c$, the expected values result
    \begin{equation}
        \begin{split}
        \left< f_{c}^{*} \right> &= \frac{c}{2} \left[ 1 + \Re \left( \left\langle \sqrt{1-\frac{n_\alpha}{n_{max}}} \right\rangle \right) \right], \\ 
        \left< e_{c}^{*} \right> &= e_{c}^{*} \cdot \left( 1-\frac{N_{o}^{e}}{N} \right),
        \end{split}
        \label{eq:EB_fe_eq_Ic}
    \end{equation}
    where $\Re(z)$ is the real part of the complex number $z$, and
    \begin{equation}
        \left\langle \sqrt{1-\frac{n_\alpha}{n_{max}}} \right\rangle = \frac{1}{M} \sum_{\alpha=1}^{M} \sqrt{1-\frac{n_\alpha}{n_{max}}}
        \label{eq:EB_nrms}
    \end{equation}
    is an expectation value.
    At the opposite limit, in scarcity conditions $q(f_\alpha)=I_sf_\alpha$, the energy expected values result
    \begin{equation}
        \begin{split}
        \left< f_{l}^{*} \right> &= f_{l}^{*} \left( \left< n \right> - \Delta^{f}_{o} \right), \\ 
        \left< e_{l}^{*} \right> &= e_{l}^{*} \left( \frac{\left< n^2 \right> - \Delta^{e}_{o}}{\left< n \right>} \right),
        \end{split}
        \label{eq:EB_fe_eq_Is}
    \end{equation}
    where the functions $f_{l}^{*}$ and $e_{l}^{*}$ are defined in Eq. \eqref{eq:EB_Is_eq}, and
    \begin{equation}
        \begin{split}
        \left\langle n \right\rangle   &= \frac{N}{M}, \\
        \left\langle n^2 \right\rangle &= \frac{\mathbf{n}\cdot \mathbf{n}}{M}, \\
        \Delta^{f}_{o}                 &= \frac{N_{o}^{f}-M_{o}^{f}n_{max}^{f}}{M}, \\
        \Delta^{e}_{o}                 &= \frac{\mathbf{n}_{o}^{e} \cdot \mathbf{n}_{o}^{e} - N_{o}^{e}n_{max}^{e}}{M},    
        \end{split}
        \label{eq:EB_dif}
    \end{equation}
    being $\left\langle n \right\rangle$ the average number of agents per patch, $\left\langle n^2 \right\rangle$ the second moment of the agent density, $\Delta^{f}_{o}$ the mean overpopulation per patch, and $\Delta^{e}_{o}$ the second moment of the overpopulation.
    Note that $\left\langle n^2 \right\rangle$, $\Delta^{f}_{o}$ and $\Delta^{e}_{o}$ grow with the system's {spatial} heterogeneity. 
    Thus, from Eq. \eqref{eq:EB_fe_eq_Is} it can be deduced that, under conditions of scarcity, {group} behaviors (highly heterogeneous) are detrimental to the environment, but may be beneficial to the species in question.

    To find the values in Eq. \eqref{eq:EB_dif} and estimate the system energies under scarcity conditions, it is necessary to know the probability $P(n)$ that a patch contains $n$ agents {in the stationary regime}. Let us recall that   
    \begin{equation}
        \begin{split}
        \Delta^{f}_{o}     ~ &= \sum_{n=n_{max}^{f}}^{\infty} P(n) \left( n - n_{max}^{f}   \right), \\
        \Delta^{e}_{o}     ~ &= \sum_{n=n_{max}^{e}}^{\infty} P(n) \left( n - n_{max}^{e}   \right) n.
        \end{split}
        \label{eq:DP_Pn}
    \end{equation}
    If the exchange of agents between patches happens faster than the relaxation time of the energies,
    the system behaves as if the patches contain a continuous number of agents, not necessarily integer. 
    Thus, when considering the {\textit{high motility}} limit ($D\rightarrow\infty$),     
    the discrete summation in Eq. \eqref{eq:DP_Pn} must be replaced by a continuous integral.
    In this regime, the probability results in
    \begin{equation}
        P_{\infty}(n) = \delta \left( \langle n\rangle - n \right), 
        \label{eq:DP_Pv}
    \end{equation}
    a highly homogeneous system.
    Therefore
    $\left\langle n^2 \right\rangle=\left\langle n \right\rangle^2$, 
    $\Delta^{f}_{o} = \left( \left\langle n \right\rangle - n^{f}_{o} \right) \Theta\left(\left\langle n \right\rangle - n^{f}_{o}\right)$ and 
    $\Delta^{e}_{o} = \left( \left\langle n \right\rangle - n^{e}_{o} \right) \left\langle n \right\rangle \Theta\left(\left\langle n \right\rangle - n^{e}_{o}\right)$, where $\Theta$ is the Heaviside step function.
    Thus, it follows
    $\left< f^{*} \right> = f^{*}(\left\langle n \right\rangle)$ and 
    $\left< e^{*} \right> = e^{*}(\left\langle n \right\rangle)$.
    {Recall that, for a constant population, $\langle n \rangle = N_0/M$, but it may change when birth and death processes are introduced.}
    
    In the \textit{static} limit $D\rightarrow0$,
    there is no effective exchange of agents between patches, so the distribution of agents remains constant. If the initial positions of the agents are random, the probability {is given by a Poisson point process,}
    \begin{equation}
        P_{0}(N;n) =  \binom{N}{n} \left( \frac{1}{M} \right)^{n}
              \left( 1 - \frac{1}{M} \right)^{N-n}, 
        \label{eq:DP_Pd}
    \end{equation}
    which is the binomial distribution corresponding to randomly dropping $N$ particles into $M$ patches.
    From this expression, it can be found
    $\left< n^2 \right>=N/M[1+(N-1)/M]$.
    Note that other initial distributions will yield different probabilities, depending on the specific problem to be modeled.

\subsection{Equilibrium Population}
\label{s:EP}

    Due to birth and death processes, the agent population evolves over time. 
    From Eq. \eqref{eq:EB_Is_eq}, the reproduction, $n_r$, and the starvation, $n_s$, populations per patch, defined as the populations at which the mean energy per agent results $e_{l}^{*} = e_r$ and $e_{l}^{*} = e_s$ respectively, are    
    \begin{equation}
        \begin{split}
        n_r &= n_{max}^{e} \left( 1 - \frac{e_r}{\varepsilon} \right), \\
        n_s &= n_{max}^{e} \left( 1 - \frac{e_s}{\varepsilon} \right).
        \end{split}
        \label{eq:EP_nrns}
    \end{equation}    
    In patches where $n_\alpha \leq n_r$, agents reproduce, whereas in patches where $n_\alpha \geq n_s$, agents die.    
    Therefore, in equilibrium, the population in each populated patch must satisfy
    \begin{equation}
        n_r < n^{*}_\alpha < n_s.
        \label{eq:EP_nb}    
    \end{equation}
    In the {high motility} limit ($D\rightarrow\infty$), the rapid exchange of agents between patches leads to a homogeneous agent density. Thus, the total system population $N_{\infty}^{*}$ results
    \begin{equation}
        n_r < \frac{N_{\infty}^{*}}{M} < n_s.
        \label{eq:EP_Nb}    
    \end{equation}
    Note that, although the number of agents is bounded, it is not determined, suggesting that $N_{\infty}^{*}$ may be sensitive to initial conditions.

    In the static limit ($D\rightarrow0$), the equilibrium population per patch $n_{0}^{*}$ can be calculated as follows: Assume a random initial distribution of $N_0$ agents in $M$ patches, with an average population per patch $n_0=N_0/M$.
    If the population $n_\alpha$ of the $\alpha$-th patch satisfies $n_\alpha \leq n_r$, agents will reproduce, increasing the population to $n_\alpha=\lceil n_r \rceil$, i.e., the smallest integer greater than $n_r$. On the other hand, if $n_\alpha \geq n_s$, agents will starve, reducing the population to $n_\alpha=\lfloor n_s \rfloor$, i.e., the largest integer less than $n_s$.
    Considering this, the equilibrium population is given by
    \begin{equation}
        n_{0}^{*} = n_0 + \lceil n_r \rceil p_< - n_< + \lfloor n_s \rfloor p_> - n_>.
        \label{eq:EP_Nd}
    \end{equation}
    where $p_<$ is the probability of a patch to have an initial population below the reproductive threshold $n_r$ and $n_<$ is the average agent population in those patches. Similarly, $p_>$ is the probability of a patch with initial population above the starvation threshold $n_s$ and $n_>$ is the average agent population in those patches. In terms of the probability distribution,
    \begin{equation}
        \begin{split}
        p_< &= \sum_{n \leq n_r} P_{0}(N_0;n), \\
        n_< &= \sum_{n \leq n_r} P_{0}(N_0;n)n, \\
        p_> &= \sum_{n \geq n_s} P_{0}(N_0;n), \\
        n_> &= \sum_{n \geq n_s} P_{0}(N_0;n)n.    
        \end{split}
        \label{eq:EP_MNrs}
    \end{equation}    
    Note that this applies only when the initial agent positions are randomly assigned in a static limit $D \rightarrow 0$.

    We previously analyzed the equilibrium population in the static and high-motility limits.
    However, the population size in the intermediate regime is non-trivial and difficult to determine.
    Nonetheless, we can identify the limiting parameters that ensure a sustainable system, based on the agent dynamics.
    Solving the agents' energy equation, Eq. \eqref{eq:M_A}, for a constant $f_\alpha$ and initial condition $e_i(0)=e_0$, it results
    \begin{equation}
        e_i(t) = e_{eq} + \left[ e_0 - e_{eq} \right]\exp\left({-mt}\right),
        \label{eq:EP_e(t)c}
    \end{equation}
    where $e_{eq}=[q(f_\alpha)-\kappa v_0^2]/m$.
    This implies that agents' energy tends asymptotically to equilibrium in a characteristic time $1/m$.
    Considering agent 's mean square displacement
    \begin{equation}
        \left< \Delta x^2(t) \right> = 4Dt,
        \label{eq:EP_xrms}
    \end{equation}
    the \textit{limit area}
    \begin{equation}
        A_l = \frac{4D}{m}
    \end{equation}
    represents the maximum (minimum) area to which an agent must have access for itself alone in order to reach the equilibrium energy $e_s$ ($e_r$) in a time $1/m$ and die (reproduce).
    The actual area to which an agent has access only for itself is
    \begin{equation}
        A=\frac{L_p^2}{\langle n \rangle}.
        \label{eq:EP_A}
    \end{equation}
    In the limiting case where $e_{eq} = e_s $ and $A \leq A_l$, after a time $1/m$, agents cannot find new patches where they can replenish their energy and die. This case always represents a phase transition for the system, beyond which there is an absorbing phase.
    In the opposite case, where $e_{eq} = e_r $ and $A \geq A_l$, after a time $1/m$, agents
    find a new free patch and exceed the reproductive energy, duplicating. This increase in population is harmless in most systems. However, if $e_{eq} > e_r$ for one agent per patch, but $e_{eq} < e_s$ for two agents per patch, this condition can lead to extinction.     
    The average equilibrium agent density can be expressed as
    \begin{equation}
        {\langle n (e_{eq}) \rangle} = \frac{m L_p^2}{4D},
        \label{eq:EP_n}
    \end{equation}
    where it has been assumed that, in a time $1/m$, a persistent random walker has covered an area $A$, i.e., $A_l=A$.
    Regarding Eq. \eqref{eq:EP_nrns}, 
    the critical values for the model parameters, i.e. those threshold values beyond which populations vanish, are given by
    \begin{equation}
        e_{s} \leq
        \frac{I_s c}{m} \left( 1 - \frac{m L_p^2 I_s D_\theta}{2 r v_0^2} \right) - \frac{\kappa v_0^2}{m} \leq
        e_{r} .
        \label{eq:EP_D}
    \end{equation}
    Note that Eq. \eqref{eq:EP_n} can not be used directly to determine the population size. 
    Eq. \eqref{eq:EP_A} determines the available area per agent, while Eq. \eqref{eq:EP_xrms} determines the minimum (maximum) consumption area that an agent must have available in order not to die (reproduce).    
    When there is a large population, it is not correct to assume that areas in Eqs. \eqref{eq:EP_A} and \eqref{eq:EP_xrms} are identical. Therefore, Eq. \eqref{eq:EP_n} is only valid near the extinction limit.

\section{Numerical Results}
\label{NR}

    Analyzing the influence of agents and environmental characteristics on population dynamics provides insight into how these factors may favor specific movement strategies.
    {To analyze the outcome of the model introduced above under different regimes, we performed numerical simulations considering the set of parameters described in Appendix \ref{App:Par}.}

\subsection{Motion influence}
\label{s:CP}     
    
    \begin{figure*}
    \centering
    \begin{tabular}{c c c}
        \adjustbox{valign=t}{\includegraphics[width=4.3cm]{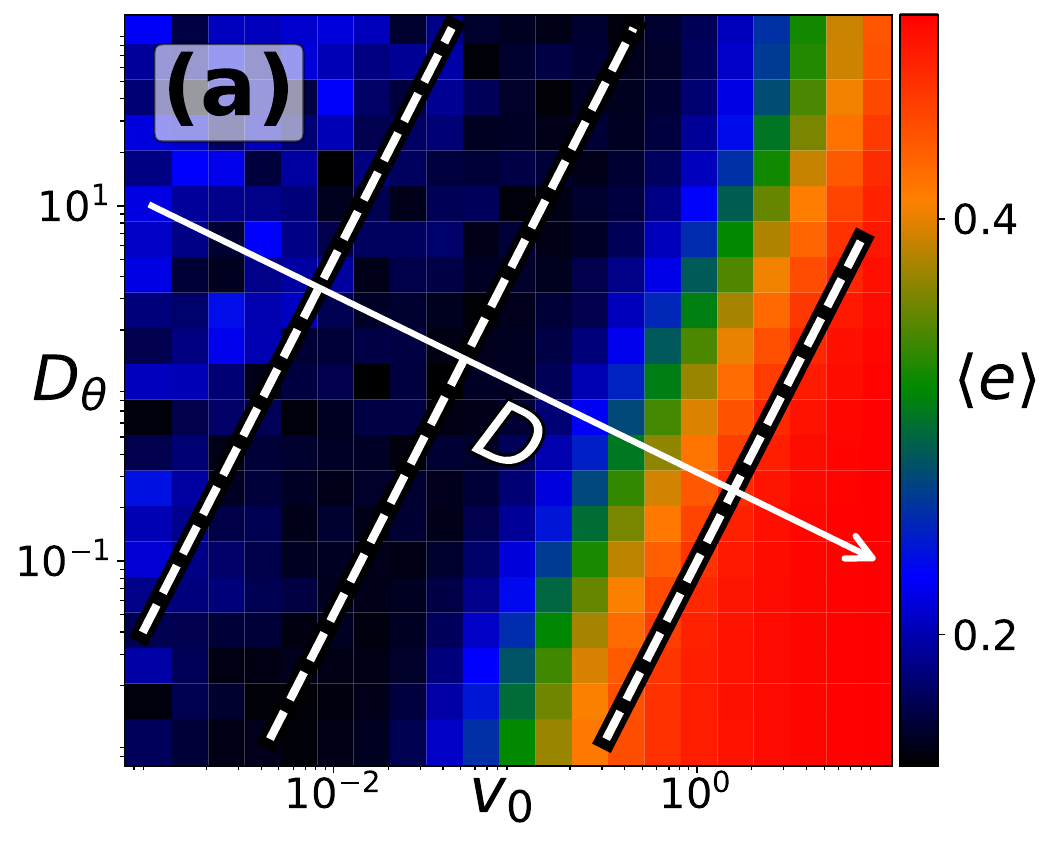}} &
        \adjustbox{valign=t}{\includegraphics[width=4.3cm]{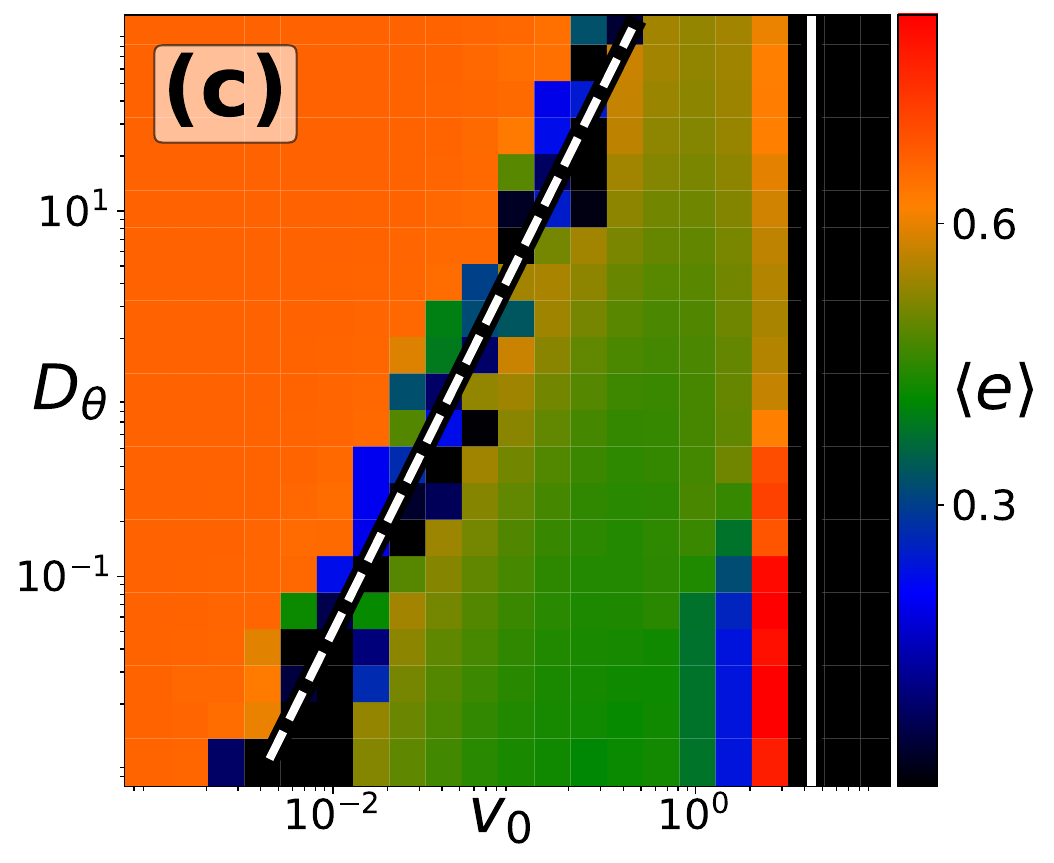}} &
        \multirow{2}{*}{\adjustbox{valign=t}{\includegraphics[width=8.6cm]{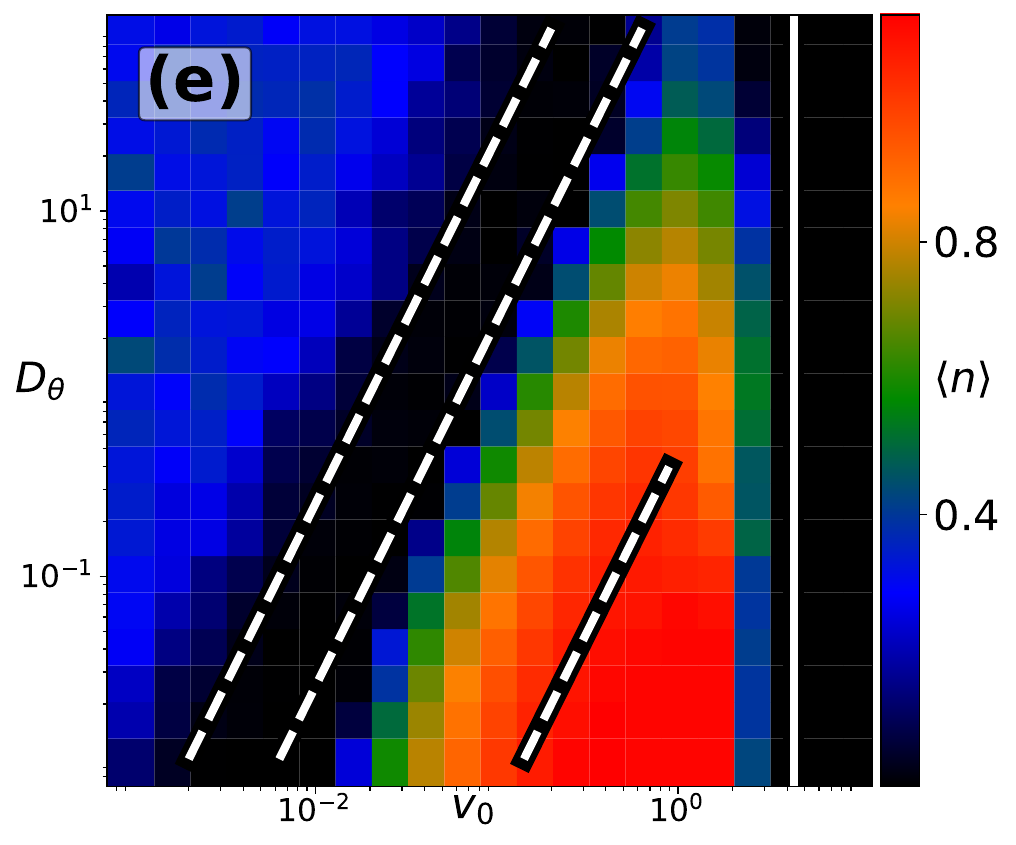}}} \\        
        \adjustbox{valign=t}{\includegraphics[width=4.3cm]{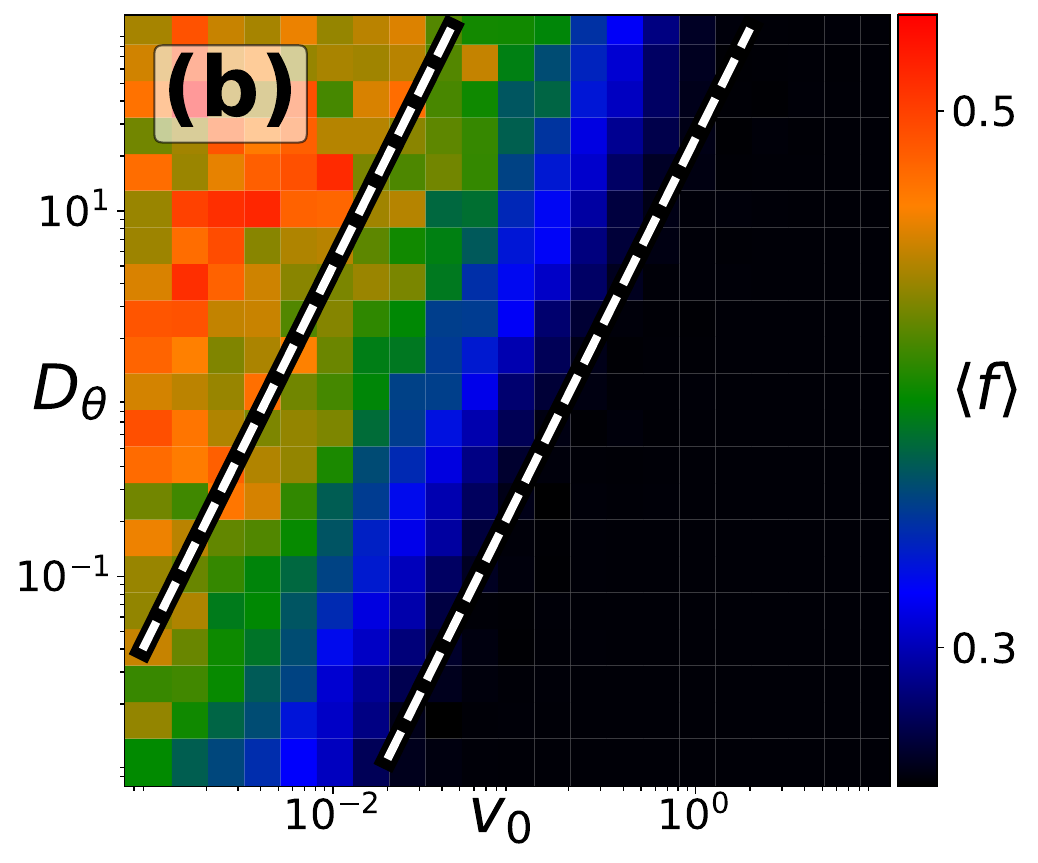}} &
        \adjustbox{valign=t}{\includegraphics[width=4.3cm]{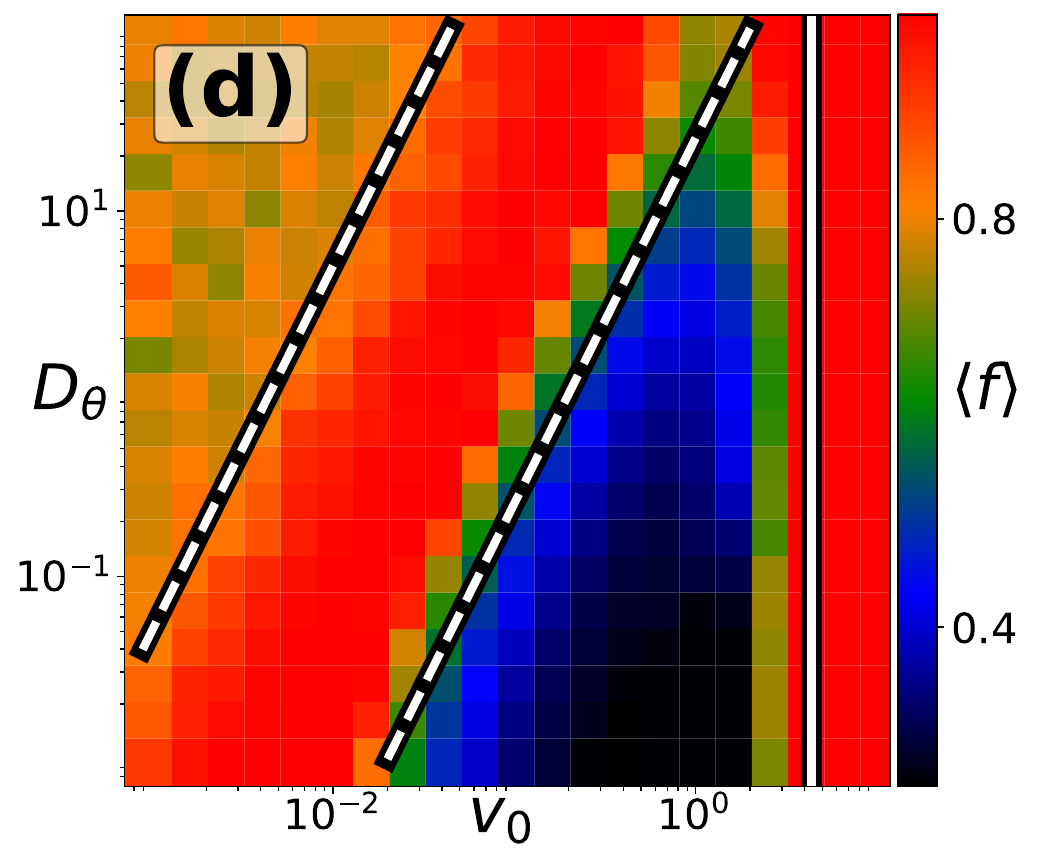}} &
    \end{tabular}
        \caption{ {Phase diagramas for: }
        \textbf{Left column}:
        \textit{Constant population},
        {$N_0=253$}, $\kappa =0$.
        \textbf{(a)} Average energy per agent $\left\langle e \right\rangle$. 
        White diagonal arrow indicates the $D$ increasing direction (same for all graphs), while dashed lines correspond to $D=10^{-5}$, $10^{-3}$ and $5$.
        Agent energy remains constant at both low and high $D$ ($\left\langle e \right\rangle \approx 0.2$ for $D<10^{-5}$ and $\left\langle e \right\rangle \approx 0.5$ for $D>5$), reaching its minimum at $D\approx10^{-3}$ ($\left\langle e \right\rangle \approx 0.13$).      
        \textbf{(b)} Average energy per patch $\left\langle f \right\rangle$.
        Diagonal dashed lines correspond to $D=10^{-5}$ and $2\cdot10^{-2}$.  
        Patch energy remains constant at both low and high $D$ ($\left\langle f \right\rangle \approx 0.45$ for $D<10^{-5}$ and  $\left\langle f \right\rangle \approx 0.25$ for $D>2\cdot10^{-2}$), and decreases monotonically with $D$ for intermediate values.   
        \textbf{Center and right columns}:
        \textit{Births and deaths}, {$N(t)$}.      
        \textbf{(c)} Average inner energy per agent {(diagonal line $D=10^{-3}$)}, 
        \textbf{(d)} Average inner energy per patch {(diagonal lines $D=10^{-5}$ and $D=2\cdot10^{-2}$)},
        \textbf{(e)} Average population per patch {(diagonal lines $D=2\cdot10^{-4}$, $D=2\cdot10^{-3}$ and $D=1$)}.
        The graphs are divided by a central diagonal band corresponding to an absorbing phase, i.e., extinction of the agent population, where high fluctuations on $\langle e \rangle$ can be observed. 
        This band separates the static regime (low $v_0$, high $D_\theta$), where $\langle e \rangle$ is high, $\langle f \rangle$ is medium and $\langle n \rangle$ is medium, from the {high motility} regime (high $v_0$, low $D_\theta$), with medium $\langle e \rangle$, low $\langle f \rangle$ and high $\langle n \rangle$.
        The emergence of a new absorbing phase for $v_{0} \geq v_{0,max}$ (vertical solid line, Eq. \eqref{eq:KR_vmax}) can be observed.
        }
        \label{fig:efn}
    \end{figure*} 

    The aim of this section is to analyze the system's energy and population equilibrium and its relation with the agents' motion, characterized by the active velocity $v_0$ and the angular diffusion $D_\theta$.      
    As a first approach, births, deaths and kinetic cost will be neglected, i.e., population is constant and $\kappa=0$.  
    
    Results are presented as phase diagrams for $N(t)=N_0$ in Fig. \ref{fig:efn} (a) and (b), where the mean values of the energies per agent, $\left\langle e \right\rangle$, in (a), and per patch, $\left\langle f \right\rangle$, in (b), are shown as functions of $v_0$ and $D_\theta$. 
    Both figures are characterized by diagonal bands, corresponding to a constant diffusion coefficient $D=v_0^2/2D_\theta$, which increases from top left to bottom right, perpendicular to the bands. This implies that the equilibrium energies do not depend on $v_0$ or $D_\theta$ separately; rather, they depend on their relationship through $D$.
    Thus, we will continue our analysis considering $D$, depicted in Fig. 1(a) as diagonal dashed lines.

    As previously stated, for $D\rightarrow 0$ the exchange of agents between patches is negligible and its distribution is obtained from Eq. \eqref{eq:DP_Pd}. Under these conditions, both $\left\langle n^2 \right\rangle$ and $\Delta_{o}^{e}$ are maximized, minimizing $[\left< n^2 \right>-\Delta^{e}_{o}]$. In Fig. \ref{fig:efn} (a) this can be observed as a noisy plateau ($\left\langle e \right\rangle \approx 0.2$) for $D<10^{-5}$, given by a binomial distribution.
    As $D$ grows, the exchange of agents between patches increases and both $\left\langle n^2 \right\rangle$ and $\Delta^{e}_{o}$ decrease monotonically. As a result, agent distribution becomes more homogeneous, leading to the disappearance of overpopulated patches, i.e. $\Delta^{e}_{o}=0$, while population variations persist, i.e. $\left\langle n^2 \right\rangle > \left\langle n \right\rangle ^2$. At this point, the argument in Eq. \eqref{eq:EB_fe_eq_Is} reaches its maximum, resulting in the minimum value for the agent energy, $\left\langle e \right\rangle \approx 0.13$, observed in Fig. \ref{fig:efn} (a) at $D=10^{-3}$.    
    Finally, as stated in Eq. \eqref{eq:DP_Pv}, for $D\rightarrow\infty$ all patches contain $\langle n \rangle$ agents, implying $\left\langle n^2 \right\rangle = \left\langle n \right\rangle ^2$ and $\Delta^{e}_{o}=0$. This is in accordance with the results in Fig. \ref{fig:efn} (a), where $\left\langle e \right\rangle = 0.5$ for $D > 5$, corresponding to evaluating Eq. \eqref{eq:EB_Is_eq} at $\langle n \rangle$, i.e. $e_{l}^{*}(\langle n \rangle)$. 

    The same analysis is valid when analyzing the average patch food. However, note that $\left\langle f \right\rangle$ depends only on $\Delta^{f}_{o}$, but not on $\left\langle n^2 \right\rangle$. Thus, $\Delta^{f}_{o}$, and therefore $\left\langle f \right\rangle$, will decrease monotonically with $D$, as can be observed in Fig. \ref{fig:efn} (b). 
    For $D < 10^{-5}$, there is a noisy plateau around $\left\langle f \right\rangle \approx 0.45$, corresponding to the binomial distribution of Eq. \eqref{eq:DP_Pd}. 
    As $D$ grows from $10^{-5}$ to $2\cdot10^{-2}$, $\left\langle f \right\rangle$ decreases monotonically. 
    For $D > 2\cdot10^{-2}$, $\Delta^{f}_{o}$ goes to zero, and $\left\langle f \right\rangle = f_{l}^{*}\left( \langle n \rangle \right) \approx 0.25$ is constant, satisfying Eq. \eqref{eq:DP_Pv}, as can be observed in Fig. \ref{fig:efn} (b).
    Therefore, the analytical results match quantitatively with the numerical results in the static and {high motility} limits, and allow a qualitative description of the intermediate results.   
    
    The average time an agent spends in a patch can be estimated from Eq. \eqref{eq:EP_xrms} as $t_p = L_p^2/4D$.
    For $D<10^{-5}$, it follows $t_p>2.5\cdot10^6$, which is a time longer than the total simulation time, so the system can be considered frozen, explaining the static behavior observed.
    For $D>5$, it results $t_p<5$, which is several orders of magnitude shorter than the relaxation time of energies for agents and patches, leading to a high exchange of agents and a fast homogenization of the system.
    In the intermediate regime, the exchange and distribution of agents between patches will have a significant impact on the energy of the system.     
     {As a consequence}, the $v_0$-$D_\theta$ plane is divided into three phases, a static phase for $D<10^{-5}$, a diffusive phase for $10^{-5}<D<5$ and a {high motility} phase for $D>5$.


    To analyze the evolution of the agent population and its relation to agent motility, from now on, we consider birth and death processes and take all the parameters described in the Appendix \ref{App:Par}. Therefore, agent population $N(t)$ will be a function of time.    
    The results are presented as phase diagrams in the $D_\theta$-$v_0$ space.    
    Fig. \ref{fig:efn} (c) shows the mean values of the energies per agent $\left\langle e \right\rangle$, Fig. \ref{fig:efn} (d) the mean values of the energies per patch $\left\langle f \right\rangle$ and, Fig. \ref{fig:efn} (e) the mean population per patch $\left\langle n \right\rangle$.   
    Even though the number of agents, and consequently $\langle n \rangle$, may vary, the analysis developed in Sec. \ref{EB} remains applicable, provided that no patches are overpopulated. The diagonal white dashed lines correspond to constant $D$ values as before. From Eq. \eqref{eq:EP_nrns}, it can be stated that $\langle n \rangle$ decreases with $\kappa v_0^2$. This effect is negligible for low $v_0$, but considerably reduces the equilibrium population for high $v_0$, leading to system extinction.
    {The maximum velocity for an agent, at which its energy balance is no longer positive,}
    \begin{equation}
    \begin{split}
        v_{0,max} = \sqrt{ \frac{I_sc - m e_s}{\kappa } },
        \label{eq:KR_vmax}
    \end{split}
    \end{equation}
    can also be deduced from Eq. \eqref{eq:EP_nrns}, by taking $n_s=0$. For the simulation parameters considered here, it results in $v_{0,max} = 4.36$, which is shown as a vertical solid white line in (c), (d) and (e).
    {We emphasize that, below the maximum speed $v_{0,max}$, the outcome of the system does not depend on $v_0$ and $D_\theta$ independently, but rather through $D$.}
    
    In the {high motility} limit, for $D>1$, $n_{\infty}^{*} = 1.15$ is obtained from the simulations, which coincides with Eq. \eqref{eq:EP_Nb} as the analytical bounds are $n_r=0.825$ and $n_s=1.425$.
    The mean population decreases with $D$ until it reaches an absorbing phase $\langle n \rangle=0$ from $D=2\cdot10^{-3}$ to $D=2\cdot10^{-4}$.
    As $D$ keeps decreasing, $n$ grows until the static limit is reached at $D=10^{-4}$.
    In this regime, it results $n_r=0.825<1$. This means that patches with populations below $n_r$ can only have $n=0$ agents, being not possible a replenishment of the population.
    At the same time, for the parameters considered, it has been observed in the simulations that patches with $n>n_s$ usually have no survivors after the resource depletion.
    This behavior is observed in nature among arvicoline rodents populations, where overshooting the carrying capacity may lead to population crashes \cite{turchin2001availability}. 
    Consequently, Eq. \eqref{eq:EP_Nd} can be reduced to   
    \begin{equation}
        n_{0}^{*} = n_i - n_>.
        \label{eq:VP_N0}
    \end{equation}
    Given the parameters of the simulations, it results $n_{0}^{*} = 0.37$, which agrees with the results observed in Fig. \ref{fig:efn} (e). In this way, the simulations validate the analytical predictions. 
    
    Due to agents' death, overpopulated patches are no longer stable, as the energies in Eq. \eqref{eq:EB_Is_eq} would be negative. Therefore, the terms $\Delta_{o}^{f}$ and $\Delta_{o}^{e}$ in Eq. \eqref{eq:EB_fe_eq_Is} vanish. As a result, $\left\langle f \right\rangle = f_{l}^{*} \left( \left\langle n \right\rangle \right)$ is satisfied for all $D$, implying that a linear increase (decrease) in $\left\langle f \right\rangle$ results in a linear decrease (increase) in $\left\langle n \right\rangle$, which is confirmed by Fig. \ref{fig:efn} (d) and (e). Furthermore, $\left\langle n^2 \right\rangle $ can be deduced from $\left\langle e \right\rangle $ (Eq. \eqref{eq:EB_fe_eq_Is} with $\Delta_{o}^{e}=0$), since $\left\langle e \right\rangle = e_{l}^{*} \left( \left\langle n^2 \right\rangle / \left\langle n \right\rangle \right)$. This highlights the importance of spatial organization, characterized by $\left\langle n^2 \right\rangle$.  
    For low $D$, the exchange of agents between patches is so small that agents in overpopulated patches cannot escape and die. If $D=0$, agent exchange is null and there are only extinctions in the initially overpopulated patches.
    As $D$ increases, agent exchange leads to spontaneous overpopulations in initially non-overpopulated patches, leading to a decrease in total population. This effect is maximal for $D \approx 10^{-3}$, resulting in an absorbing phase (extinction). 
    For $D>10^{-3}$, on the other hand, agent exchange occurs faster, allowing some of them to escape the overpopulated patches and survive. Under this condition, total population increases with $D$.    

    Note that both $n_{\infty}^{*}$ and $n_{0}^{*}$ depend on initial conditions.
    In Sec. \ref{s:EP} it was stated that the equilibrium populations must satisfy the Eqs. \eqref{eq:EP_Nb} and \eqref{eq:EP_Nd}. However, note that Eq. \eqref{eq:EP_Nb} does not give an exact population, but a range within which the equilibrium population can lie, while Eq. \eqref{eq:EP_Nd} depends on the initial conditions.    
    In the limit $D \rightarrow \infty$, energies and distribution are highly homogeneous, evolving the total population to a certain value in the range $[n_r,n_s]$, as can be observed in Fig. \ref{fig:IP}.  
    In the opposite limit, $D \rightarrow 0$, agents are static and the final population strongly depends on the initial distribution. In general, agent number in overpopulated patches will decrease to $n_s$ while population in underpopulated patches will grow to $n_r$.
    On the other hand, far from these limits, the population is much more dynamic; its equilibrium value fluctuates widely and does not depend on the initial conditions (see Fig. \ref{fig:IP} for $D=10$).
    Thus, two dynamics locally maximize the agent population: a static strategy with low $D$ and a {high motility} strategy with high $D$, separated by an absorbing phase. 
    Intermediate mobility emerges as a maladaptive strategy as it maximizes extinction risk.
    The contribution of the kinetic rate is to constrain the {high motility} strategy, therefore, it is bounded by absorbing phases at both low $D$ and high $v_0$.    
    {Comparing Figs. \ref{fig:efn} (c) and (e), it is noteworthy that, in the static limit, although the energy per agent is high, the population is low. The opposite occurs in the {high motility} limit, where the population is high but the energy per agent is low. 
    
    {In nature, it is expected that populations will match one of these strategies.
    There could be species that maximize individual intake ({selfish behavior}) while others maximize the population {(group behavior)}}
    \cite{goodhall2002altruism}. 

    \begin{figure*}
        \centering
        \includegraphics[width=4.2cm,trim=0 0 0 0]{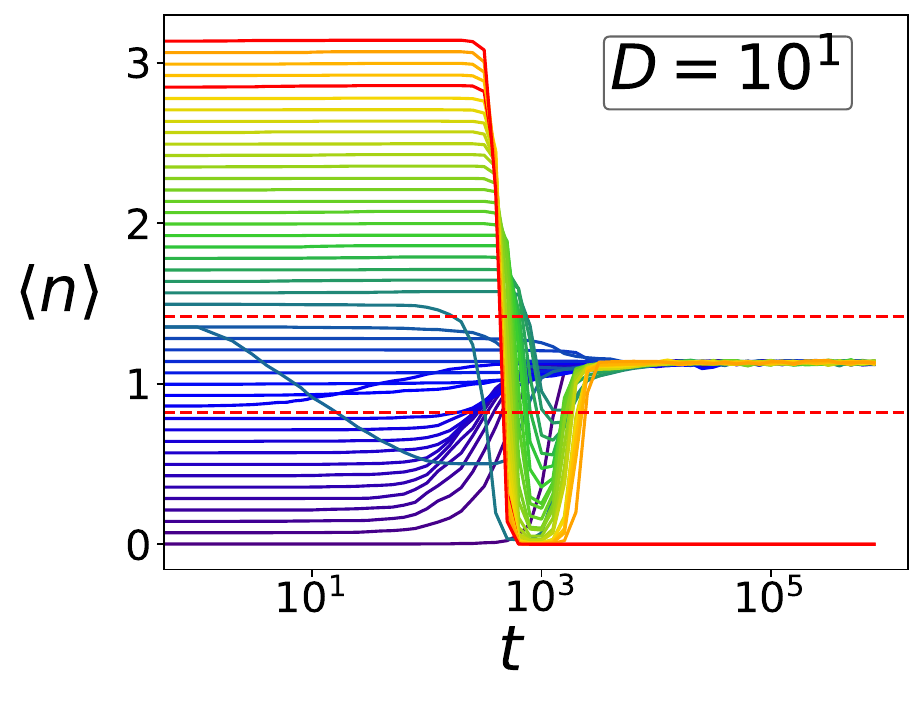}
        \includegraphics[width=4.2cm,trim=0 0 0 0]{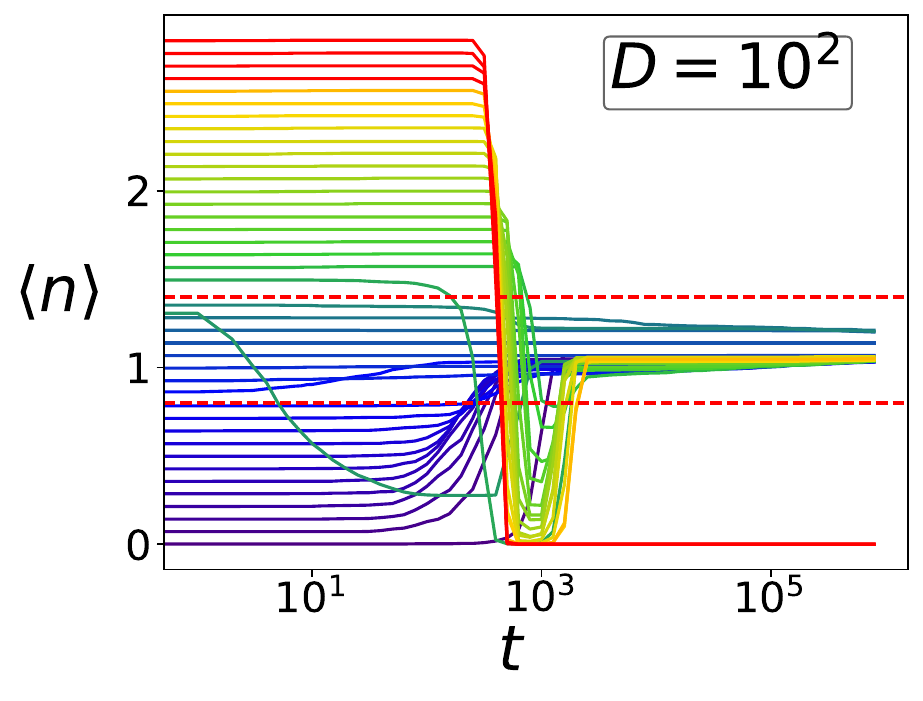}
        \includegraphics[width=4.2cm,trim=0 0 0 0]{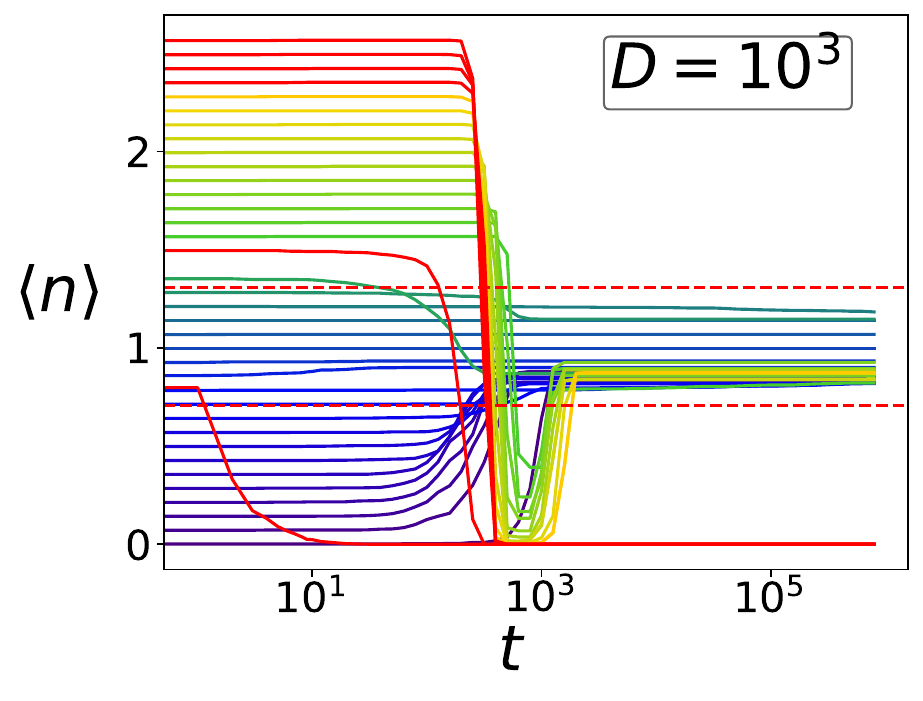}
        \includegraphics[width=4.2cm,trim=0 0 0 0]{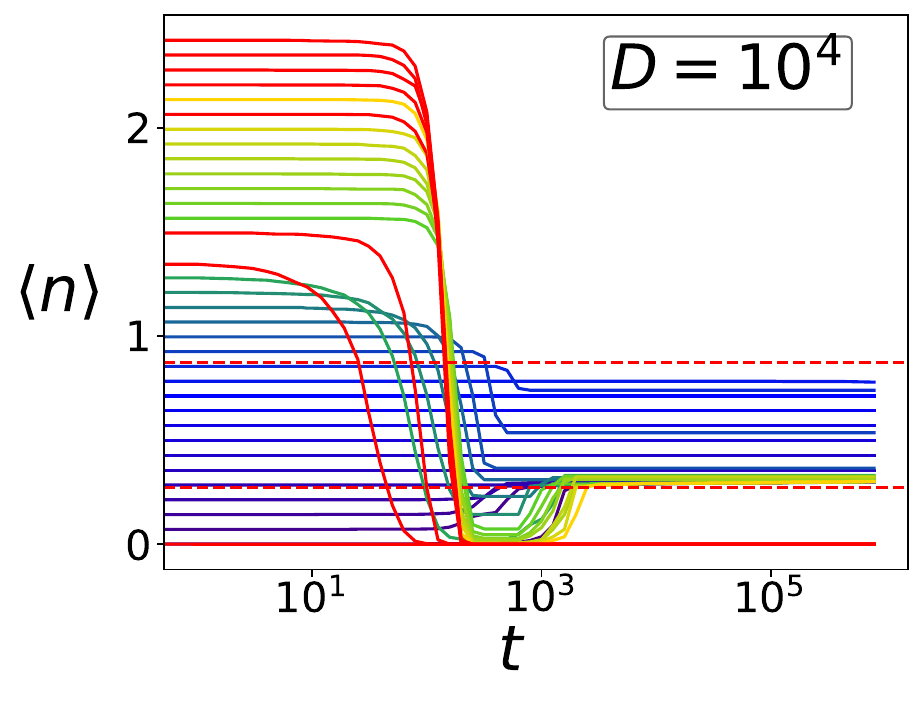}
        \includegraphics[width=4.2cm,trim=0 0 0 0]{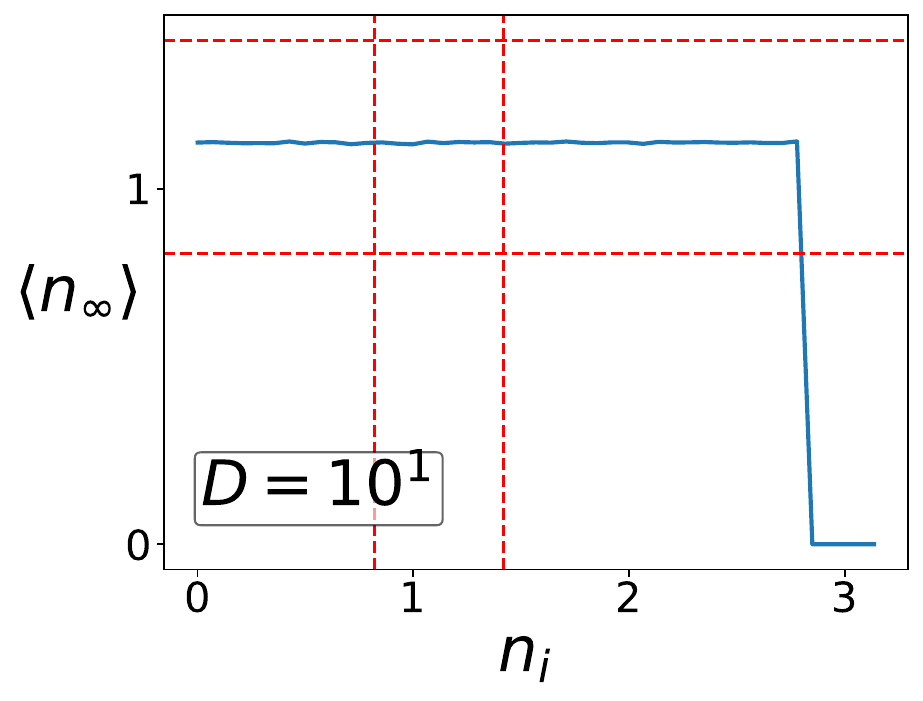}
        \includegraphics[width=4.2cm,trim=0 0 0 0]{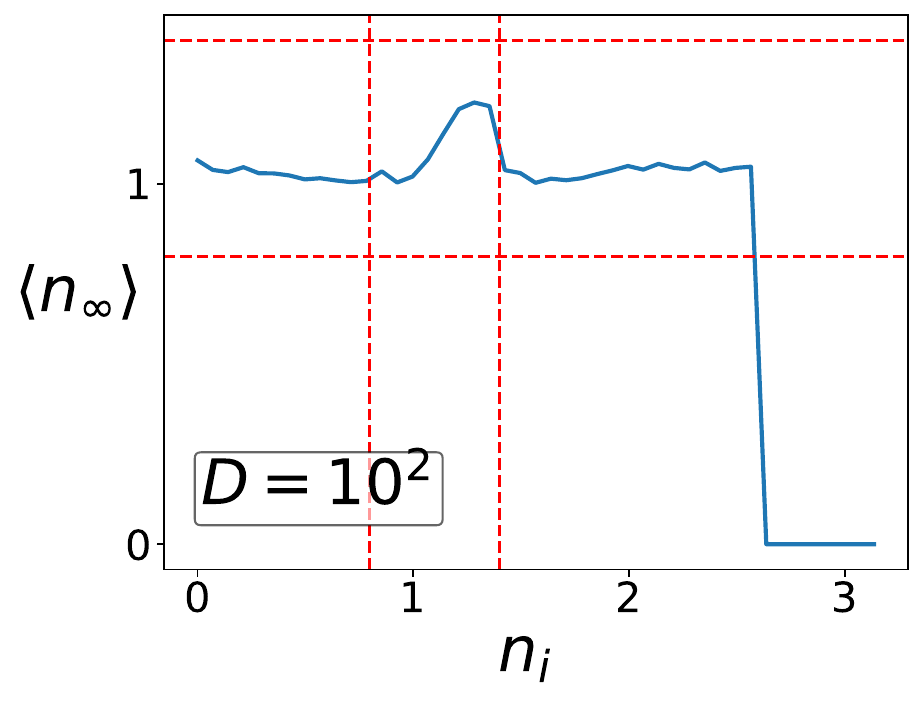}
        \includegraphics[width=4.2cm,trim=0 0 0 0]{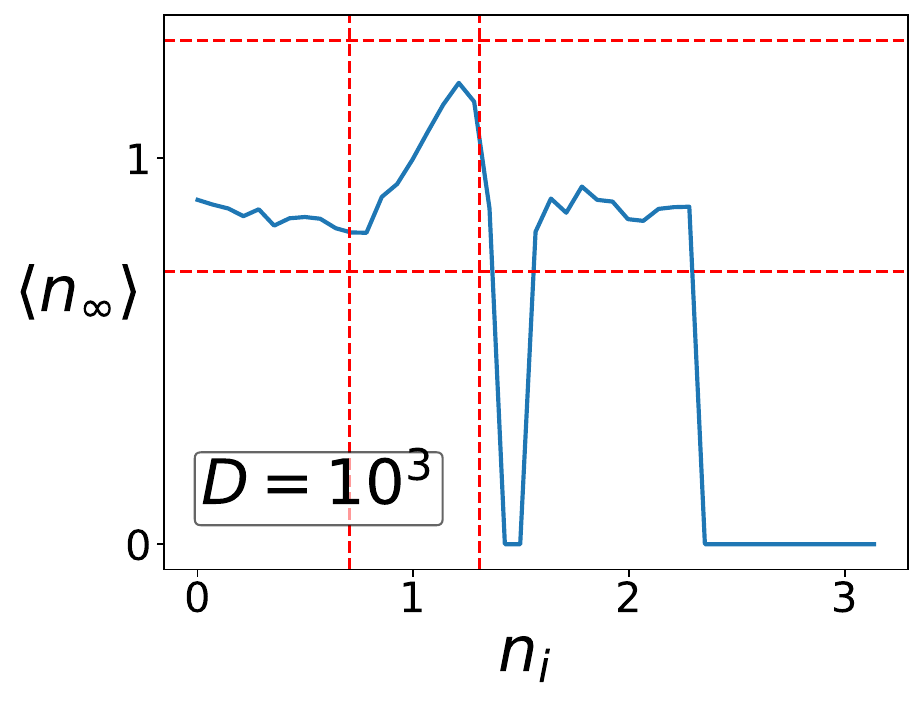}
        \includegraphics[width=4.2cm,trim=0 0 0 0]{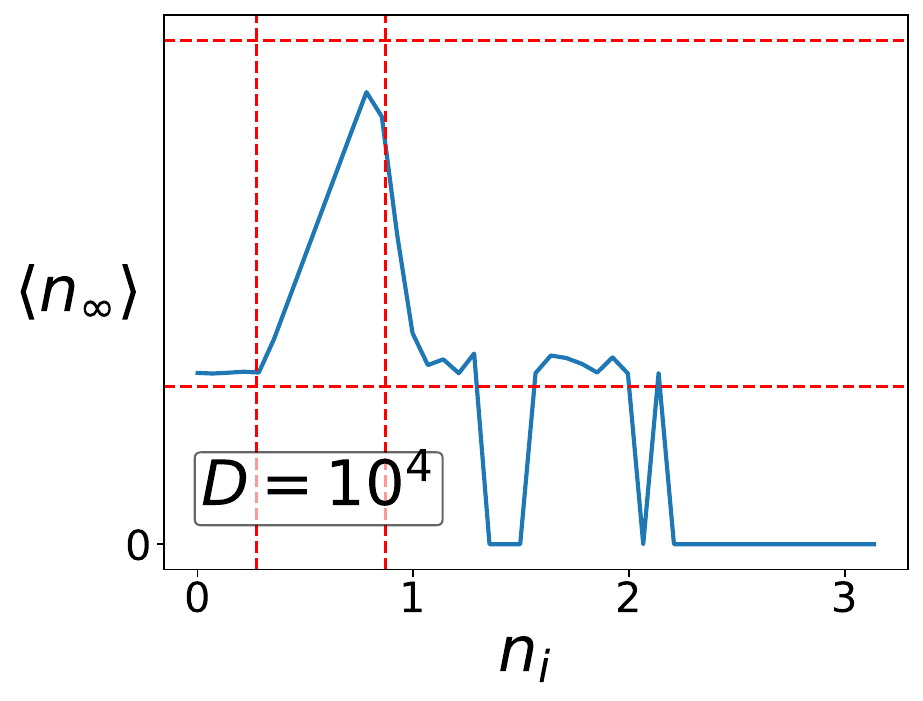}
        \caption{
        \textbf{Upper row}: Time evolution of the agent populations for different initial populations and diffusion coefficients {(one realization per line)}.
        \textbf{Bottom row}: Equilibrium population as a function of the initial population for different diffusion coefficients.
        Red dashed lines correspond to $n_r$ and $n_s$.
        For $D\leq 10$, the equilibrium population does not depend on the initial conditions. 
        For $D>10$, the equilibrium population depends on the initial population, presenting a linear relationship in the region $n_r<n_i<n_s$ and large fluctuations for $n>n_s$. However, the equilibrium population is always bounded within the range $[n_r,n_s]$, when it does not become extinct.
        }
        \label{fig:IP}
    \end{figure*} 
    
    Figure \ref{fig:IP} shows the time evolution of population size and its dependence on the initial condition, $n_i$, for the {high motility} regime. The dashed red lines indicate the values of the reproductive, $n_r$, and starvation, $n_s$, populations, derived from Eq. \eqref{eq:EP_nrns}. 
    As can be seen, unless an extinction occurs, the equilibrium population always lies within this range, i.e., $\langle n_\infty \rangle \in \left[ n_r, n_s \right]$.
    It can be inferred that, for a given initial population, there is an associated extinction probability that increases with $n_i$. 
    These extinctions take place in systems that are below the maximum speed $v_{0,max}$, and therefore have a well-defined stationary state. Nevertheless, during the transient time, extreme oscillations in the population occur, leading to extinction in finite systems (see Appendix \ref{App:Osc})}.    
    Similar oscillations are observed in a tritrophic model which includes vegetation, rodents, and predators \cite{turchin2001availability}.    
    For systems with $D \leq 10$, the equilibrium population is independent of $n_i$, with $n_\infty\approx(n_r+n_s)/2$. 
    For higher values of $D$ the behavior is more complicated.
    On the one hand, if the initial population is outside the range $[n_r,n_s]$, the equilibrium population will approach $n_r$ as $D$ increases.
    On the other hand, if $n_i \in [n_r,n_s]$, populations will show little variation over time, with these variations decreasing as $D$ increases. 
    
    We highlight that, although the system is highly dependent on the initial conditions, the analytical limits correctly bound the equilibrium populations.

\subsection{Energy parameters} 

\begin{figure*}
        \centering
        \includegraphics[width=5.5cm,trim=0 0 0 0]{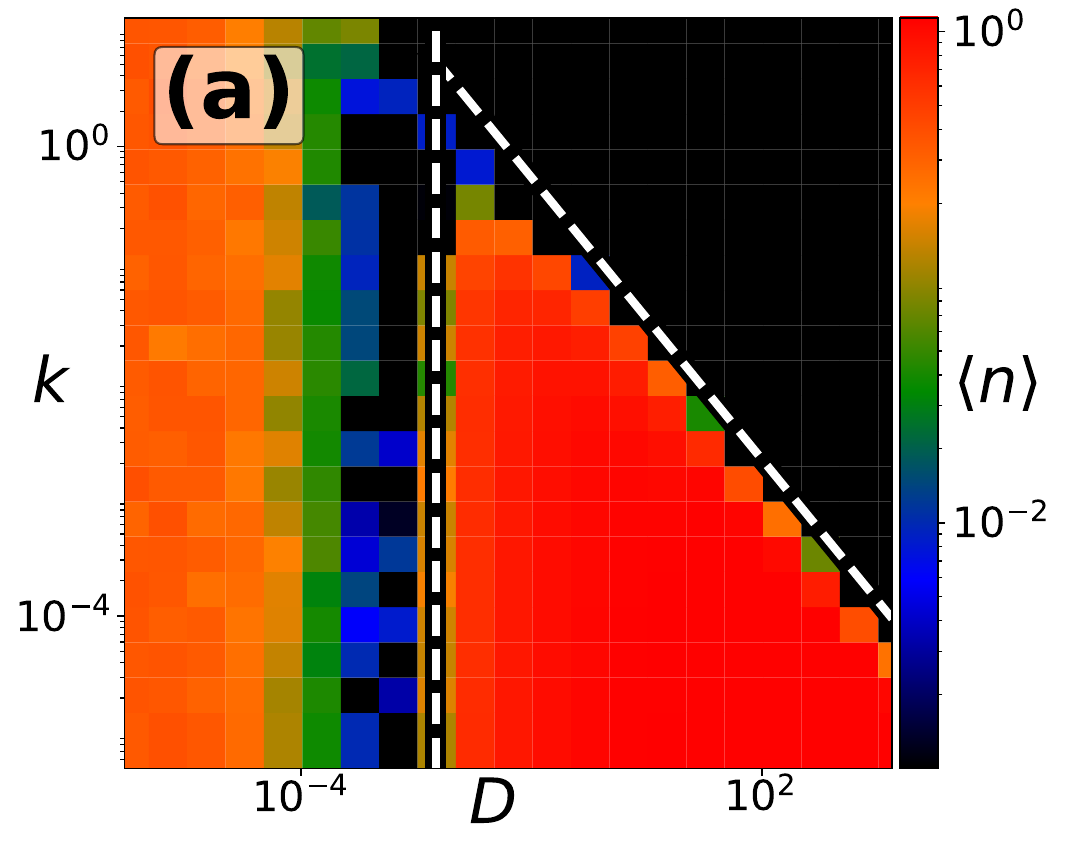}
        \includegraphics[width=5.5cm,trim=0 0 0 0]{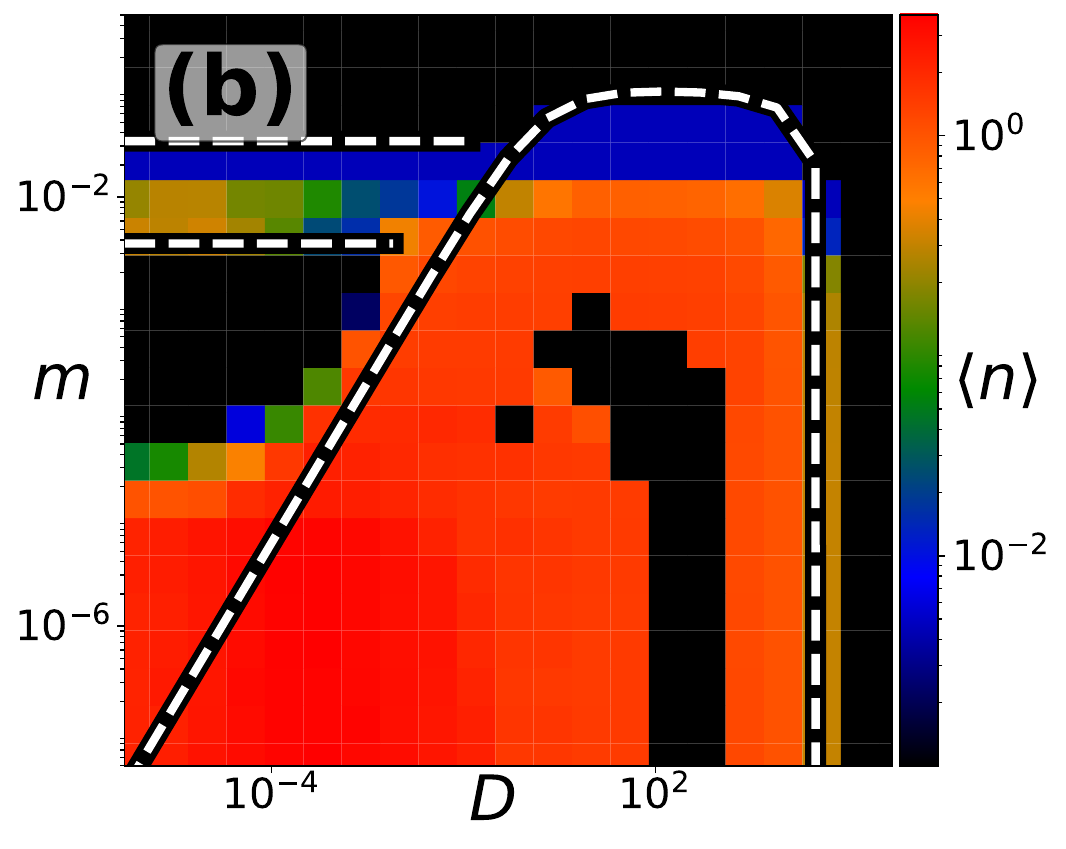}
        \includegraphics[width=5.5cm,trim=0 0 0 0]{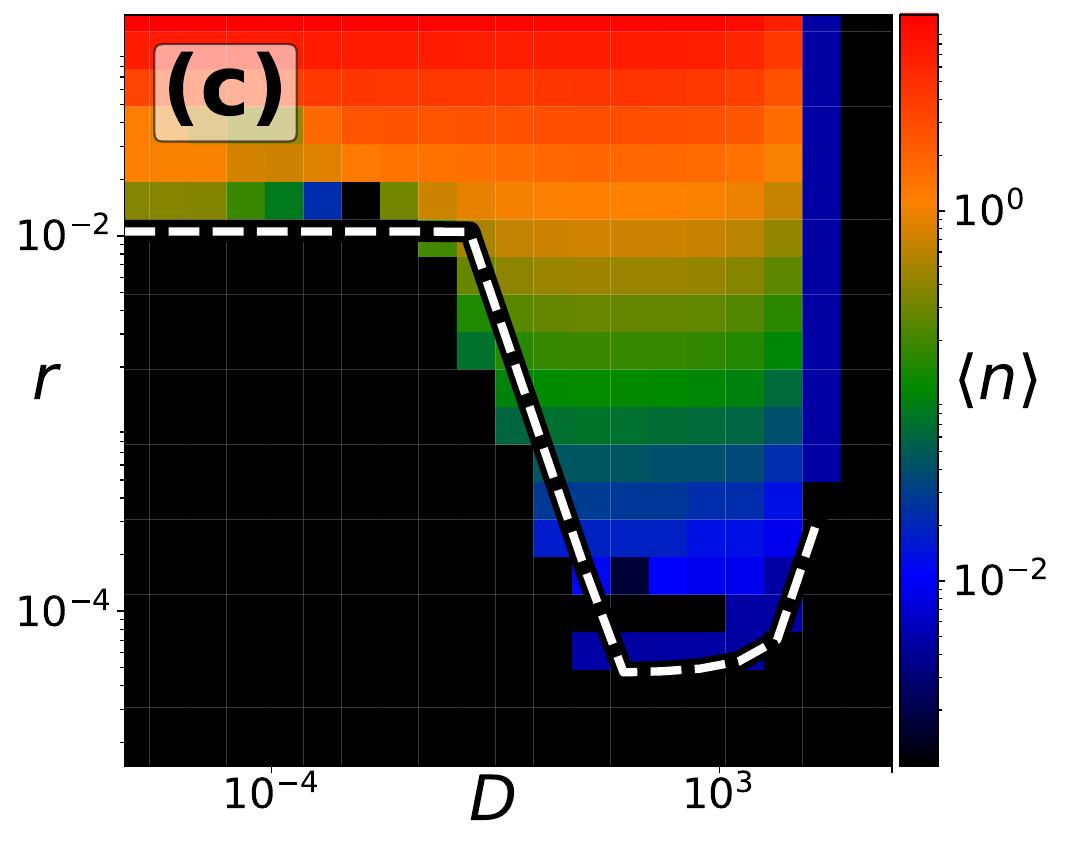}
        \includegraphics[width=5.5cm,trim=0 0 0 0]{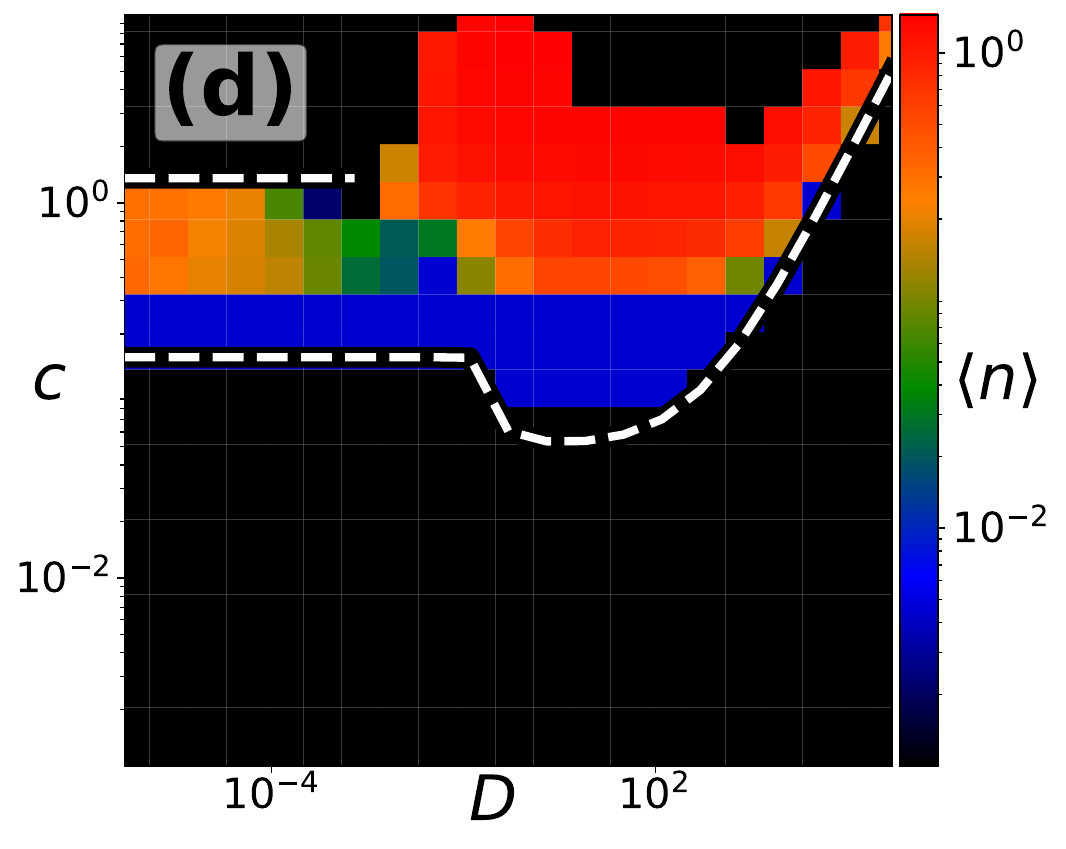}
        \includegraphics[width=5.5cm,trim=0 0 0 0]{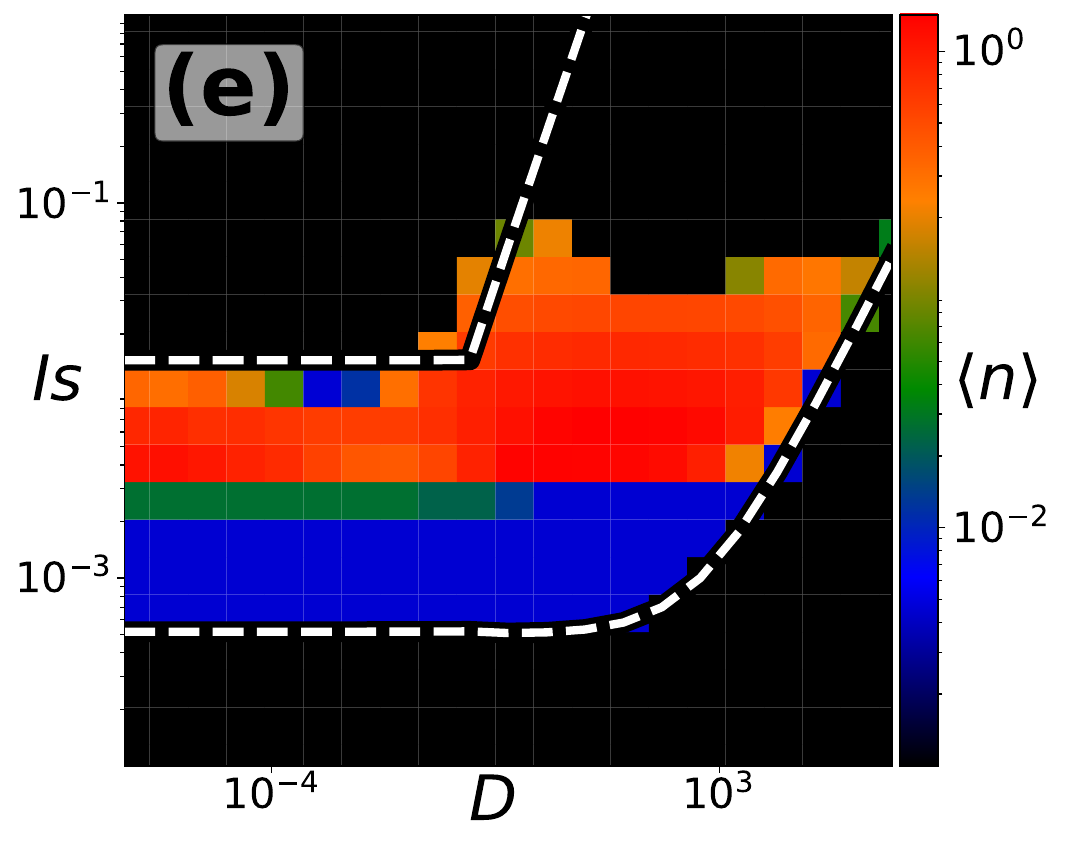}
        \includegraphics[width=5.5cm,trim=0 0 0 0]{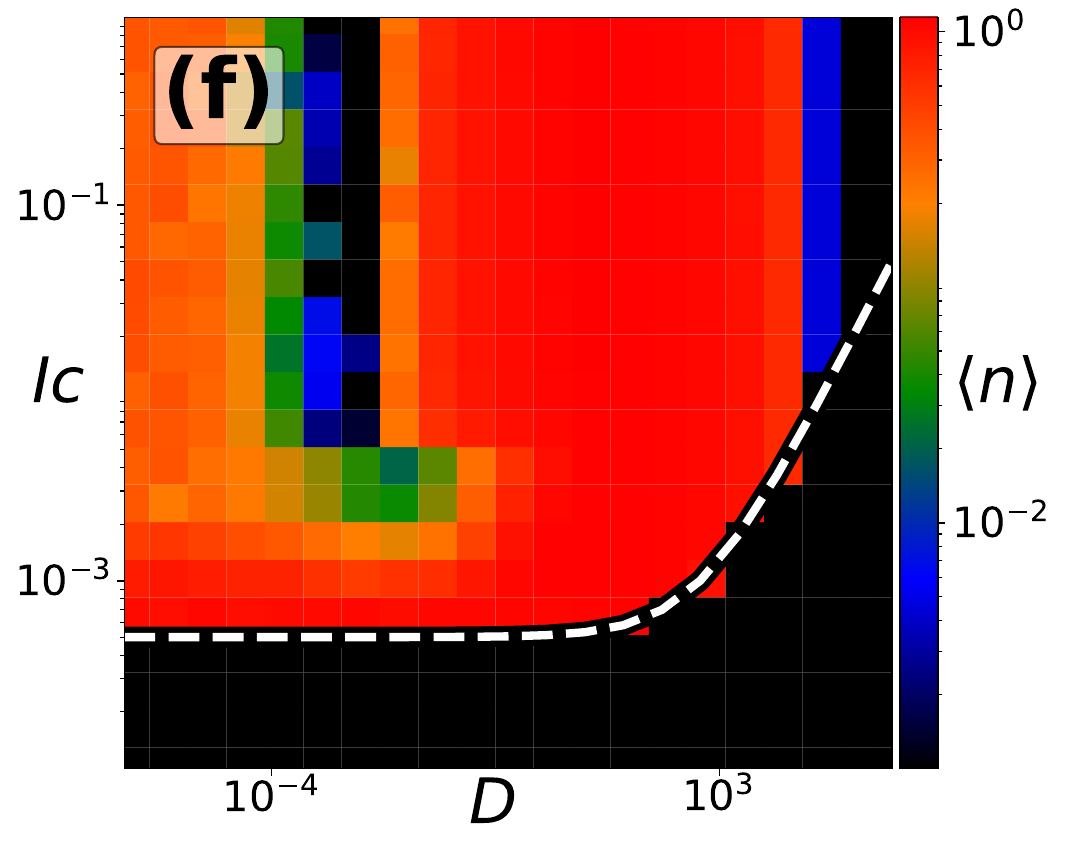}
        \caption{     
        Heat map of agents equilibrium population $\langle n \rangle$ as a function of diffusion coefficient $D$ and    
        \textbf{(a)} kinetic rate $\kappa$, 
        \textbf{(b)} metabolic rate $m$, 
        \textbf{(c)} growth rate $r$, 
        \textbf{(d)} patch capacity $c$, 
        \textbf{(e)} intake slope $I_s$, 
        \textbf{(f)} intake capacity $I_c$.    
        \textbf{(a)} \textit{kinetic rate}.
        White dashed vertical line corresponds to Eq. \eqref{eq:EP_D},
        white dashed diagonal line corresponds to Eq. \eqref{eq:KR_k}.   
        The figure shows two absorbing phases corresponding to these Eqs. 
        \textbf{(b)} \textit{metabolic rate}.
        Main white dashed line corresponds to Eq. \eqref{eq:MR_m1}, while top and bottom horizontal dashed white lines correspond to Eq. \eqref{eq:MR_m}. These lines delimit the absorbing phases of the system.
        \textbf{(c)} \textit{growth rate}.
        White dashed line corresponds to Eq. \eqref{eq:GR_r}, indicating the phase transition.
        \textbf{(d)} \textit{patch capacity}.
        Main white dashed line corresponds to Eq. \eqref{eq:PC_c}, while top left white dashed line corresponds to $n_r=1$. Both lines delimit the boundary between the active phase and the absorbing phases.
        \textbf{(e)} \textit{intake slope}.
        White dashed line corresponds to Eq. \eqref{eq:Is}, separating the active phase of the system from the absorbing phases.
        \textbf{(f)} \textit{intake capacity}. 
        White dashed line corresponds to Eq. \eqref{eq:Ic_Ic}.
        }
        \label{fig:parameters}
    \end{figure*}

    In this section, we will explore the impact of different energy parameters of the model in the population size of the system.    
    Fig. \ref{fig:parameters} shows the average population per patch $\langle n \rangle$ as a function of the diffusion coefficient $D$ and (a) the kinetic rate $\kappa$, (b) the metabolic rate $m$, (c) the growth rate $r$, (d) the patch capacity $c$, (e) the intake slope $I_s$ and (f) the intake capacity $I_c$.
    {In order to cover a larger range in $D$, $v_0$ and $D_\theta$ were varied at the same time (seting $D_\theta \sim v_0^{-1}$), resulting in $D\sim v_0^3$.}


    The kinetic rate indicates how much energy it costs the agents to move. 
    It is to be expected that species with a high kinetic rate avoid spending resources on active movement. 
    On the other hand, species with a low kinetic rate have no problem spending resources on self-propulsion, which offers them the opportunity to actively explore the environment in search of more resources.
    Fig. \ref{fig:parameters} (a) shows a heat map of $\langle n \rangle$ as a function of $\kappa$ and $D$.  
    The population shows two habitable regions, separated in {high motility} (high $D$) and static (low $D$) regimes. 
    The upper bound for the {highly motile} population can be obtained from Eq. \eqref{eq:KR_vmax}, resulting
    \begin{equation}
        \kappa_{ub} = \frac{I_sc-me_s}{v_0^2}.
        \label{eq:KR_k}
    \end{equation}
    For small kinetic rate, $ \kappa \leq 0.3 $, the maximum population in the {high motility} regime, $D_{b}$, is larger than the population in the static regime, $D_{s}$. However, as $ \kappa$ grows, the {high motility} maximum population decreases until it completely disappears, being the static strategy the only available. 
    Therefore, there is a critical value, $\kappa_c $, for which the optimal agent dynamics changes from {high motility} to static, and consequently the system presents a phase transition.     
    Those species for which moving through the environment involves a considerable expenditure of energy (quantified by a high value of the kinetic rate $ \kappa $), 
    {will have to reduce their active velocity $v_0$ to maintain their energetic balance, thus, reducing their diffusion coefficient $D$ into the static regime.}
    On the contrary, species for whom moving through the environment does not represent a great energetic cost (low value of the kinetic rate $ \kappa $), will take advantage of the energy available in the environment by adopting a {high motility} strategy. 
    Finally, if the kinetic rate of the agents is close to the critical value $ \kappa_{c} $, both behaviors can coexist.


    Different species have different efficiencies in terms of resource management. This can be modeled by the metabolic rate, which indicates how much energy an agent requires to fulfill its metabolic functions. Figure \ref{fig:parameters} (b) shows $\langle n \rangle$ as a function of $m$ and $D$. There, the main white dashed line corresponds to
    \begin{equation}
        m_{ub} = \frac{r}{I_s} \left( 1 - \frac{\kappa v_0^2}{I_sc} \right) \left( \frac{L_p^2}{4D} + \frac{re_s}{I_s^2c} \right)^{-1},
        \label{eq:MR_m1}
    \end{equation}
    which can be derived from Eq. \eqref{eq:EP_D}, and describes the boundary between the absorbing phase and viable systems. 
    Although simulations generally agree with Eq. \eqref{eq:MR_m1}, they reveal a richer and more complex behavior. The available area per agent, estimated in Eq. \eqref{eq:EP_A}, indicates the number of patches that an agent feeds on in a time $1/m$. Therefore, it does not make sense if this area is smaller than $L_p^2$, the area of a single patch (for nearly static agents), or bigger than $ML_p^2$, the area of the system (for {highly motile} agents).
    The top and bottom horizontal dashed white lines correspond to 
    \begin{equation}
        m_{s/r} = \frac{1}{e_{s/r}} \left( I_sc -\frac{I_s^2c}{r} \kappa v_0^2 \right)
        \label{eq:MR_m}
    \end{equation}
    which can be derived from Eq. \eqref{eq:EP_nrns} by taking $n_s=1$ for $e_s$ and $n_r=1$ for $e_r$ respectively, defining the boundaries of a new habitable zone. This corresponds to the static regime observed previously.
    Above the line corresponding to $n_s=1$, the metabolic rate is so high that a patch is not able to sustain a single agent, so the population starves. Below the line corresponding to $n_r=1$, a patch provides enough food for a single agent to reproduce, but not enough to support two agents. Thus, the population will experience an increase followed by a collapse, leading to extinction. 
    Note that this behavior is due to finite size effects.    
    We can conclude that, as metabolic cost increases, sedentary strategies become less efficient, making it necessary to explore the environment in search of resources by adopting wandering strategies.
    Similar behaviors could be observed in ungulate species \cite{keeping2014animal} and in models of fish larvae \cite{darowski1988bioenergetic}.
    Finally, at the center of the {high motility} region, an absorbing phase appears due to time oscillations in the populations, which can lead to extinctions because of the finite size of the system (see Appendix \ref{App:Osc}). 
    

    Different ecosystems have different recovery and replenishment rates. In our model, they are represented through the growth rate, $r$, which indicates the rate at which the patches produce resources. Fig. \ref{fig:parameters} (c) shows a heat map of $\langle n \rangle$ as a function of $r$ and $D$. The white dashed line corresponds to the boundary between the absorbing phase and the active phase. 
    The diagonal segment follows a power law $r \sim D^{-1}$ that can be derived from Eq. \eqref{eq:EP_D} as
    \begin{equation}
        r_{lb} = I_s \left( 1 - \frac{me_s+\kappa v_0^2}{I_sc} \right)^{-1} \langle n \rangle,
        \label{eq:GR_r}
    \end{equation}
    where $\langle n \rangle$ corresponds to Eq. \eqref{eq:EP_n}.
    However, this power law is truncated at both low and high $D$. For small enough $D$, we must again consider that the minimum feeding area to which an agent has access is always equal to $L_p^2$. Therefore, the horizontal segment of the line corresponds to an equilibrium between the consumption of a single agent ($\langle n \rangle=1$) and the total nutrient replenishment of a single patch.
    The finite size of the patches then truncates the diagonal line and replaces it with a horizontal one, extending the habitable zone.
    On the other hand, on the right side of the graph, when $D$ is large, we must now consider that the maximum feeding area to which an agent has access is the size of the system $ML_p^2$. The curved line to the right of the graph corresponds to $\langle n \rangle=1/M$. 
    Sedentary strategies ($D\rightarrow0$) are only sustainable for high $r$, and take advantage of the finite size of the patches. 
    As $r$ decreases, it becomes necessary to explore a larger environment for resources, adopting {highly motile} strategies (larger $D$). 
    Finally, if $r$ is small enough, the environment is unable to sustain any agent population.  
    On the other hand, if $r$ is large enough, there is an overabundance of resources, resulting in no optimal movement strategy. 
    
    Our results are consistent with previous individual-based models for marine bacteria \cite{sibona2007evolution} and stream salmonids \cite{railsback2002analysis},
    where the organisms have to look for new nutrient sources when conditions are scarce or adopt a static behavior when nutrients are abundant.

    

    Different environments also present different resources abundance, represented in our model by the patch capacity $c$. In Fig. \ref{fig:parameters} (d), $\langle n \rangle$ is shown as a function of $c$ and $D$. 
    The dashed white line marks the boundary beyond which the system cannot sustain any population:
    \begin{equation}
        c_{s/r} = \frac{me_{s/r}+\kappa v_0^2}{I_s} \left( 1 - \frac{I_s}{r} \langle n \rangle  \right)^{-1},
        \label{eq:PC_c}
    \end{equation}
    where $\langle n \rangle$ corresponds to Eq. \eqref{eq:EP_n} considering $e_s$.
    By following the same considerations stated in the previous graphs, the critical  average population per patch is restricted to $[1/M, 1]$. 
    Note that the curve has a minimum at $D_{c,min}\approx5$, which is consistent with the analytical result derived from Eq. \eqref{eq:PC_c},
    \begin{equation}
        D_{c,min} = \frac{I_sL_p^2m}{4r}\left(1+\sqrt{1+\frac{2re_s}{I_sL_p^2D_\theta\kappa}}\right).
        \label{eq:PC_Dmin}
    \end{equation}
    Thus, in harsh conditions, only a few movement strategies will survive.   
    For large $c$, however, we see another absorbing phase. This is again due to the population oscillatory dynamic, that leads to extinction (see Appendix \ref{App:Osc}). An overabundance of resources can lead to excessive population growth. The subsequent environmental devastation may be too extreme for the population to survive.
    This is the finite size effect explained above. In an analytical model where $n$ is continuous, the population will experience an oscillatory behavior.  
    
    The white horizontal line in the upper left corner of the graph corresponds to the solution of Eq. \eqref{eq:PC_c} taking $e_r$ and $n_r=1$. 
    Again, the finite size of the patches allows the existence of a quasi-static population, i.e., low $D$.    
    For similar species coexisting in the same environment, the smaller one (with higher effective $c$) can adopt a sedentary strategy, while the larger one (with lower effective $c$), needs to be highly mobile to survive \cite{keeping2014animal}.
    In the figure, it can be seen that the population grows monotonically with $c$. This behavior can be deduced from Eq. \eqref{eq:EP_nrns}, resulting in
    \begin{equation}
    \langle n \rangle \sim n_{max}^{e} \left( 1 - \frac{m e_s}{c I_s - \kappa v_0^2} \right).        
    \end{equation}


    In the next figure (\ref{fig:parameters} (e)), the equilibrium population is shown as a function of $I_s$ and $D$. Recall that agents regulate their food intake through the intake slope $I_s$.  The  white borderlines, which separate the active from the absorbing phases of the system, are derived from Eq. \eqref{eq:EP_D} and are the solutions of
    \begin{equation}
        I_{s,ub/lb}=\frac{r \pm \sqrt{r^2 - 4 \langle n \rangle \frac{me_s+\kappa v_0^2}{c}}}{2 \langle n \rangle},
        \label{eq:Is}
    \end{equation}
    where $\langle n \rangle$ is derived from Eq. \eqref{eq:EP_n}, restricted again to the range $1/M \leq \langle n \rangle \leq 1$.
    When $I_s$ is too small, agents cannot gather enough resources to keep their metabolic expenses and die. 
    On the other hand, Eq. \eqref{eq:M_E} implies that if $I_s>r$, then $f^*_l<0$, leading to extinction.    
    However, the figure shows some populations exceeding this threshold.
    This happens because the borderline $I_s=r$ is obtained from a mean-field approximation, which considers a homogeneous system. 
    In addition, there is an absorbing phase between the two white lines for high values of $D$ and $I_s$, due to finite size effects. Here, a continuous system would present temporal oscillations in food and agent density (see Appendix \ref{App:Osc}).
    Note that a species exhibiting excessively aggressive foraging behavior may deplete available resources, thereby increasing its risk of extinction {\cite{fryxell2008multiple}}.
    The value of the intake slope that maximizes the agents' population can be calculated as
    \begin{equation}
        I_{s,max} = 2 \frac{me_r+\kappa v_0^2}{c}.
        \label{eq:Is_nmax}
    \end{equation}    
    by considering Eqs. \eqref{eq:EP_nrns} and \eqref{eq:EB_Is_n0}.
    It is noteworthy that a high consumption strategy is not optimal. Populations can grow by reducing the gathering rate. The maximum equilibrium population will result from a trade-off between high consumption rates, i.e. short-term strategies, and low consumption rates, i.e. long-term strategies of resource rationing {\cite{barta1995frequency}}.

     
    In Fig. \ref{fig:parameters} (f), $\langle n \rangle$ is studied as a function of  $D$ and $I_c$, the intake capacity, which represents the maximum amount of food an agent can obtain from a patch.
    The minimum $I_c$ necessary for the population not to become extinct can be derived from the equation \eqref{eq:M_A}, resulting
    \begin{equation}
        I_{c,lb} = me_s+\kappa v_0^2,
        \label{eq:Ic_Ic}
    \end{equation}
    which corresponds to the white dashed line in the figure.
    As can be deduced from Eq. \eqref{eq:M_q}, intake capacity becomes irrelevant when $I_c > I_sc$, since it results $q(f)\approx I_sf$. This results in no upper boundary for populations in terms of $I_c$.    
    A remarkable fact is that, as $I_c$ approaches the border, the equilibrium population in the quasi-static regime increases, disappearing the absorbing phase that separated the low and high $D$ strategies, already commented on Fig. \ref{fig:parameters} (a). In fact, for low $D$, the maximum population is obtained at the minimum $I_c$.    
    In this way, the rationing of available resources can improve population conditions, giving rise to long-term collective strategies, 
    a behavior observed in both models and experiments in human collective behavior \cite{goldstone2008emergent}. 
    
\section{Conclusions}

    In this work we present a model for mobile populations interacting with a dynamic environment, interpreted as a nutrient source, aimed at identifying optimal movement strategies driven by resource availability. We consider one species of persistent random walkers that forage resources from the environment and store them in an inner energy depot, which they use for self-propel, sustain metabolic functions, avoid starvation, and reproduce. Agents with depleted depot die, while those with excess energy reproduce. The environment is modeled as a two-dimensional surface where resources are regenerated through logistic growth. Thus, population size and resource distribution become emergent properties of the system. Although direct interactions between individuals are absent, competition arises through shared consumption of resources. This indirect coupling leads to the emergence of optimal movement strategies that maximize population viability. Despite its simplicity, the model exhibits a variety of behaviors observed in natural ecosystems.    

    Our framework integrates an agent-based model, used to perform numerical simulations, with an analytical model that provides a formal description of the system in specific limiting cases. Both approaches yield consistent results and reveal two distinct movement strategies, corresponding to local maxima of population size in the $D_{\theta}$-$v_0$ space: a static strategy, in which individuals show minimal displacement and maintain high internal energy but small populations, and a high-motility strategy, where agents move actively, storing less energy but sustaining larger populations. The suitability of each strategy depends on species traits and habitat characteristics.
    These strategies correspond to two different foraging regimes. Individuals may either move slowly, conserving energy while waiting to encounter new nutrient sources{\cite{reichhardt2014active, martin2016reconciling}, or actively navigate the environment to locate resources. Kinetic expenditure sets an upper bound on mobility: high speeds lead to excessive energy consumption and may drive the species to extinction. Thus, species with high kinetic costs are expected to avoid unnecessary movement.      

   The trade-off between population size and stored energy establishes a link between movement strategies and group-level resource access. In our model, large populations arise under high motility, while individuals with high stored energy belong to small, more static populations. Similar behaviors have been reported in both active matter models and real ecological systems \cite{parrish1999complexity, goodhall2002altruism}.
  
    It is noteworthy that, when there is a resource shortage in a constant population, the analytical results predict a strong dependence of the mean agent energy on the agent distribution.
    From Eq. \eqref{eq:EB_fe_eq_Is}, and considering that the second moment of agent density, the mean overpopulation per patch, and the second moment of this overpopulation grow with the system's heterogeneity, it can be deduced that, under conditions of scarcity, group behaviors (highly heterogeneous) are detrimental to the environment but may be beneficial to the species 
    \cite{sumpter2010collective, keeping2014animal}.   

    Our numerical simulations show that under scarce conditions (slow resource replenishment or low environmental capacity) static strategies are less competitive, being viable only because of the finite size of the patches, when they do not become unsustainable. Thus, species are forced to adopt a highly motile behavior to avoid extinction. In the other limit, for fast resource replenishment, there is no optimal strategy.  
    Moreover, as expected, an excessive limitation on resource access always leads to extinction.
    Outside this range, however, the impact of restricting resources depends on the mobility strategy. In the high motility regime, such constraints have no significant effect on population dynamics.    
    In contrast, for static agents, limiting resource extraction promotes population growth, suggesting the emergence of long-term resource rationing as an adaptive strategy, while excessive extraction rates, associated with short-term overexploitation of resources, result in population decline among these species \cite{barta1995frequency, sibona2007evolution, goldstone2008emergent}.

    Our results are qualitatively consistent with previous findings in systems as diverse as real elk ecosystems \cite{fryxell2008multiple} and fish simulations \cite{railsback2002analysis}: when food replenishment is sufficient (high $r$), organisms can remain in safe habitats with low displacement (low $D$), optimizing survival.
    When the replenishment is low (low $r$), organisms must increase movement (high $D$) to find food, facing greater risks and metabolic costs to avoid extinction.
    They also show that when resources are extremely scarce, even greater mobility cannot sustain the population.
    
    In our model, highly mobile agents show larger oscillations in population size than static agents, similar to observations in African mammals \cite{ryan2017competition}. 
    Moreover, our simulations show ``absorbing phases'' generated by extreme amplitude oscillations in population which lead to extinctions, even though the system has a non-zero theoretical equilibrium population. These extinctions occur due to the finite size of the system and the discretization of population values (see Appendix \ref{App:Osc}). 
    Similar behavior is also observed in arvicoline rodent populations. In \cite{turchin2001availability}, it is shown empirically and theoretically that in environments with extreme abundance of resources, or with abrupt increases in food availability, rapid population growth and overexploitation of the habitat can occur, followed by drastic collapses driven by overconsumption and resource supplies degradation, drastically reducing the recovery capacity of vegetation and generating boom-bust cycles.
    For elk populations \cite{fryxell2008multiple}, it was suggested that extreme resource conditions can amplify fluctuations in density and food availability, producing oscillatory dynamics and potentially collapses.
    
    Overall, we emphasize that the results presented in this work highlight the crucial role of environment-mediated interactions in shaping population dynamics. The proposed model, despite its simplicity, captures key ecological features and reproduces a wide variety of behaviors observed in nature, having applications in biology and ecology, bridging the gap between them and active matter. 
    Of course, other processes must be included to apply our model to real systems. It could be of interest to analyze how our results are modified while including processes such as chemotaxis, agent alignment, predator presence, inter- and intra-species competition or cooperation, and non homogeneous environments. Work on some of these lines is already in progress. In \cite{BSP_2025}, we extend this framework to investigate emergent collective phenomena, such as clustering and traveling waves, arising from indirect interactions through resource consumption. 
    Nevertheless, the present work provides a valuable framework for studying the impact of climate change on animal behavior and population dynamics \cite{masello2017animals, gooden2025seasonal, muhling2025climate} by modeling its influence on food sources. It can be used to predict critical habitat thresholds that may lead to species extinction, as well as to assess the effectiveness of restoration and conservation efforts. Furthermore, it provides a tool for evaluating environmental protection strategies aimed at habitat recovery, thereby enabling the development of optimized approaches to preserve biodiversity and maintain ecosystem stability.

\begin{acknowledgments}
This work was partially supported by the grants PIP 112-2015 01- 00644CO and SeCYT-UNC 05/B457.
G.B. acknowledges the Internal Doctoral Grant from CONICET and the France Excellence Eiffel Scholarship from Campus France.
\end{acknowledgments}

\appendix

\section{Agent Interaction Potential}
\label{App:Pot}

    The repulsive potential between agents is defined as
    \begin{equation}
        U(r) = \begin{cases}
            a(v_0)\left[ \left( \frac{r}{2R} \right)^{-b} - 1 \right] & \text{if } r < 2R \\
            0 & \text{if } r \geq 2R
            \end{cases},
        \label{eq:U}
    \end{equation}
    where $b$ is a constant and $a(v_0)$ is a linear function of the active velocity, $a(v_0) = c_1+c_0v_0$ such that the maximum overlapping area between two agents is independent of $v_0$.

\section{Parameters}
\label{App:Par}

    Unless explicitly stated otherwise, the set of parameters used is:
    an initial population of $N_0=M(n_s+n_r)/2$ disk-shaped agents of radius $R=1$, metabolic rate $m=0.005$ and kinetic rate $\kappa =0.0005$. The environment consist of a square area of side $L=150$, with periodic boundary conditions, divided into $M=225$ squared patches of side $L_p=10$, each with a nutrient capacity $c=1$ and nutrient growth rate $r=0.05$. Agents have an intake slope $I_s=0.01$ and an intake capacity $I_c = 0.01$. The starvation, reproduction and birth energies are $e_s=0.1$, $e_r=0.9$ and $e_b=0.4$, respectively.
    The potential parameters are $b=1$, $c_1=0$ and $c_0=1.96$.

\section{Oscillations}
\label{App:Osc}

    \begin{figure}[]
        \centering
        \includegraphics[width=8.5cm,trim=0 0 0 0]{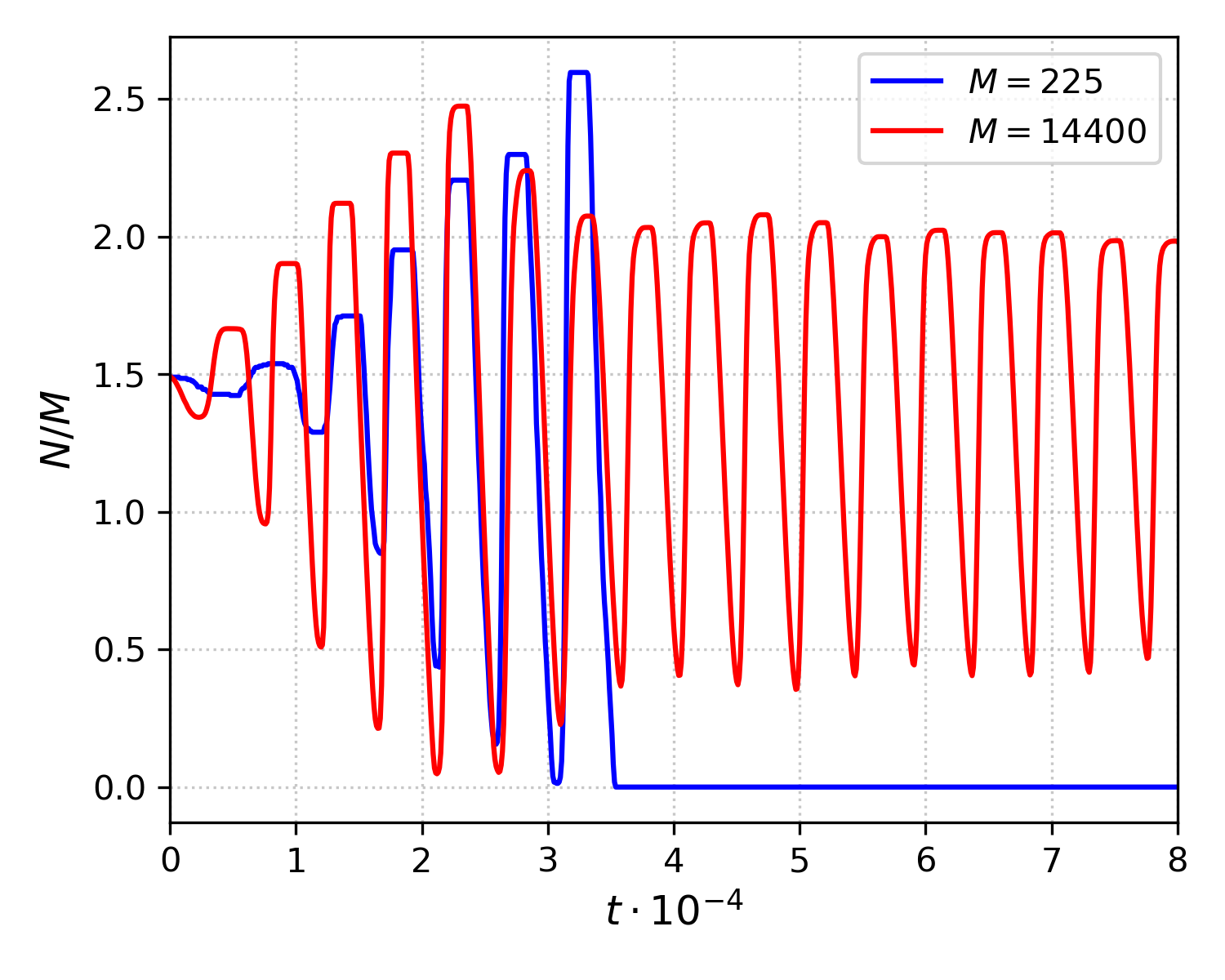}
        \caption{     
        Time series for the agent population size $N$ in systems with an unstable non-trivial equilibrium point ($D=10^2$, $m=10^{-4}$).
        For $M=225$, the amplitude grows until $N$ reaches zero and the population becomes extinct. For $M=14400$, the amplitude stabilizes after a transient period, resulting in a periodic population.
        }
        \label{fig:Osc}
    \end{figure}  
    
    For certain parameter choices, the non-trivial equilibrium size of the agent population becomes unstable, exhibiting oscillations. When these oscillations become extreme, they can lead to agent extinctions due to the finite size of the system, as can be seen in Fig. \ref{fig:Osc}. There, we show time series for the agent population size in systems with an unstable non-trivial equilibrium point ($D=10^2$, $m=10^{-4}$), for $M=225$ and $M=14400$ patches. In both systems, the population deviates from the initial equilibrium, oscillating around it with an amplitude that increases over time. It evidences the impact of the system finite size on the extinction probability. Extinction could occur for small systems, while the extinction probability is strongly reduced for large systems. 
    Although these oscillations are periodic and their frequency and amplitude can be derived from the system parameters, their study goes beyond the scope of this work.


\bibliography{bibliography}

\end{document}